\documentclass[11pt,a4paper]{article}

\pdfoutput=1
\usepackage{jheppub} 
\usepackage{subfig}
\usepackage{tabularx} 
\usepackage{gensymb}
\usepackage{soul}

\usepackage{graphicx} 
\usepackage{epstopdf} 
\usepackage{dcolumn} 
\usepackage{xcolor} 
\usepackage{amsmath,amssymb} 
\usepackage{hyperref} 
\usepackage[utf8]{inputenc} 
\usepackage[english]{babel} 
\usepackage[mathlines]{lineno} 
\usepackage{blindtext} 
\usepackage{ifthen} 
\usepackage{orcidlink} 
\usepackage{textcomp}  

\usepackage[normalem]{ulem}

\graphicspath{{figures/}} 

\newboolean{articletitles}
\setboolean{articletitles}{true} 

\input{belle2-symbols}

\input{belle2sym.tex}

\title{Search for an Axion-Like Particle in $\BtoKa$ Decays at Belle}
\collaboration{The Belle and Belle II Collaborations}
  \author{I.~Adachi\,\orcidlink{0000-0003-2287-0173},} 
  \author{L.~Aggarwal\,\orcidlink{0000-0002-0909-7537},} 
  \author{H.~Ahmed\,\orcidlink{0000-0003-3976-7498},} 
  \author{Y.~Ahn\,\orcidlink{0000-0001-6820-0576},} 
  \author{H.~Aihara\,\orcidlink{0000-0002-1907-5964},} 
  \author{N.~Akopov\,\orcidlink{0000-0002-4425-2096},} 
  \author{S.~Alghamdi\,\orcidlink{0000-0001-7609-112X},} 
  \author{M.~Alhakami\,\orcidlink{0000-0002-2234-8628},} 
  \author{A.~Aloisio\,\orcidlink{0000-0002-3883-6693},} 
  \author{N.~Althubiti\,\orcidlink{0000-0003-1513-0409},} 
% ...

\begin{document} 

\abstract{

We report a search for an axion-like particle $\ap$ in  $\B\to\kaon^{(*)}\ap$ decays using data collected with the Belle detector at the KEKB asymmetric-energy electron-positron collider. 
The search is based on a $711 \invfb$ data sample collected at the $\FourS$ resonance energy,
corresponding to a sample of $772\times10^6$~$\FourS$ events. 
In this study, we search for the decay of the axion-like particle into a pair of photons, $\ap\to\gaga$. 
We scan the two-photon invariant mass in the range 
$0.16\gev-4.50 \gev$ for the $\kaon$ modes and
$0.16\gev-4.20 \gev$ for the $\Kstar$ modes. 
No significant signal is observed in any of the modes,
and 90\% confidence level upper limits are established on the coupling to the $W$ boson, $\gaW$, as a function of $\ap$ mass. 
The limits range from $3 \times 10^{-6} \gev^{-1}$ to $3 \times 10^{-5} \gev^{-1}$, improving the current constraints on $\gaW$ by a factor of two over the most stringent previous experimental results.
}

\maketitle 
\flushbottom

 
\captionsetup[subfigure]{labelformat=empty}
\section{Introduction} 
 
The Peccei-Quinn (PQ) theory introduces axions as a solution to the strong-CP problem, positioning them as promising candidates for dark matter in extensions to the standard model (SM)~\cite{PhysRevLett.38.1440,PhysRevLett.40.223,PhysRevLett.40.279}.
Other extensions to the SM introduce axion-like particles (ALPs) $\ap$, which share the quantum numbers of axions, 
but, unlike the QCD axion, have couplings that are independent of their masses.
While the PQ axion is expected to have a mass below $\mathcal{O}(1 \mev)$ inversely proportional to its decay constant $f_a$ ($m_a \approx 6.3\,\text{eV} \cdot 10^{6}\,\gev/f_a$),
ALPs encompass a broader mass spectrum.
ALPs have the potential to address fundamental problems~\cite{Preskill:1982cy,Abbott:1982af,Dine:1982ah}. 
In particular, they can act as dark matter mediators through the axion portal at the $\mathcal{O}(\GeV)$ scale~\cite{PhysRevD.79.075008}. 

Recent years have seen a surge of interest in ALPs in the MeV and GeV mass range
~\cite{Mimasu:2014nea,Dolan:2014ska,Jaeckel:2015jla,Dobrich:2015jyk,Knapen:2016moh,Dolan:2017osp,Brivio:2017ije,Bauer:2017nlg,Choi:2017gpf,Bauer:2017ris,Mariotti:2017vtv,CidVidal:2018blh}.
The ALP couplings to photons, leptons, and gluons have been extensively studied by collider and beam dump experiments
~\cite{CHARM:1985anb,Riordan:1987aw,Bjorken:1988as,Blumlein:1990ay,OPAL:2002vhf,Aloni:2018vki,Belle-II:2020jti,LHCb:2025gbn}. 
In contrast, the coupling to $\W^\pm$ bosons remains relatively uncharted~\cite{PhysRevLett.118.111802}.
The most recent study of ALP coupling to $\W^\pm$ bosons was conducted by the \babar{} experiment using the  $\Bp\ra\Kp\ap(\ra\gaga)$ decay~\cite{BaBar:2021ich}. 
In this paper, we work in natural units ($\hbar = c = 1$), and charge-conjugate modes are implicitly included.
Although the ALP is assumed to primarily couple to the $W$ boson, gauge-boson mixing induces a coupling to photons, resulting in a nearly 100\% branching fraction for the $\ap\to\gaga$ decay in the analysed mass range, $\ma < M_W$. 

\begin{figure}[t]
\centering
\includegraphics[width=0.75\linewidth]{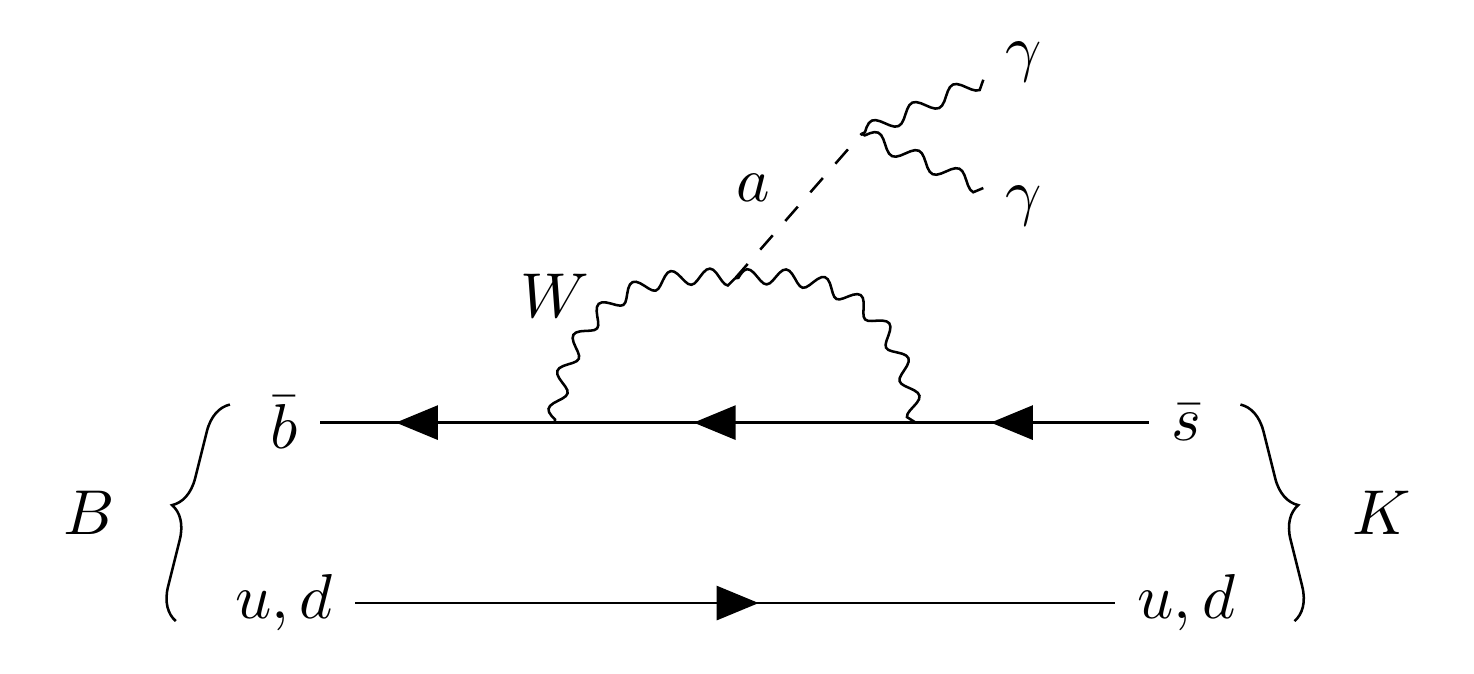}
\caption{\label{fig:feynman} The Feynman diagram for a $\BtoKa$ decay.}
\end{figure}

The coupling of the ALP to $W^{\pm}$ bosons is described by the Lagrangian~\cite{PhysRevLett.118.111802},
\begin{equation}\begin{aligned}
\mathcal{L} = -\frac{\gaW}{4} a W_{\mu\nu} \tilde{W}^{\mu\nu},
\label{eq:Lagrangian}
\end{aligned}   \end{equation}
where $a$ is the ALP field,
$\gaW$ is the coupling strength of the $\ap$ to $W$ bosons, and
$W_{\mu\nu}$ is the gauge boson field strength for a $W$ boson,
with dual tensor $\tilde{W}^{\mu\nu} = \epsilon^{\mu\nu\rho\sigma}W_{\rho\sigma}/2$, 
where $\epsilon^{\mu\nu\rho\sigma}$ is the four-dimensional Levi-Civita symbol. 
The branching fraction of the process $\BtoKa$, which is shown in Fig.~\ref{fig:feynman}, depends quadratically on  $\gaW$ (see the details  in the Appendix~\ref{sec:A}), 
reflecting the fact that the ALP decay width is given by 
\begin{equation}
\Gamma= \frac{1}{\tau_a} = \gaW^2\ma^3\sin^4\theta_W/64\pi~,
\label{eq:Gamma}
\end{equation}
where $\theta_W$ is the weak mixing angle, and $\tau_a$ is the ALP lifetime~\cite{BaBar:2021ich}.

In this paper, we report a search for ALPs  in  $\BtoKa$ decays using four kaon modes, $\KS$, $\Kp$, $\Kstarz$, and $\Kstarp$. 
The data were collected with the Belle detector~\cite{Belle:2000cnh} at the asymmetric-energy $e^+e^-$ KEKB collider~\cite{Kurokawa:2001nw}, operating at a centre-of-mass energy of $10.58\gev$, 
corresponding to the $\FourS$ resonance in $\epem$ collisions.
This search is based on a data set of $772 \pm 11$ million $\Upsilon(4S)$ mesons, corresponding to an integrated luminosity of $711$~$\invfb$. 
Our study has better sensitivity than the previous \babar{} study,
leveraging the higher total integrated luminosity of the Belle experiment, as well as additional kaon modes.
The ALP mass hypotheses 
for the $\Kp$ and $\KS$ modes range from $0.16\gev$ to $4.50\gev$, while for the $\Kstarp$ and $\Kstarz$ modes the range is $0.16\gev$ to $4.20\gev$.
This analysis also probes regions of parameter space in which the ALP lifetime is
non-negligible, leading to displaced decays a few centimetres  from the interaction point.
Once we obtain the results separately for each mode, 
we combine the four kaon modes using a simultaneous fit to improve the constraint on $\gaW$.

\section{The Belle detector and simulation}

The Belle detector is a general purpose detector, 
described in detail in~\cite{Belle:2000cnh}.
It has a cylindrical symmetry around the beam line,  with the $z$-axis being defined as the direction opposite to the positron beam. 
The detector consists of six subdetectors:
a silicon vertex detector (SVD) for precise vertex determination, 
a central drift chamber (CDC) for reconstruction of charged particle trajectories (tracks) and for measuring their momentum, 
an aerogel Cherenkov counter (ACC) and 
a time-of-flight scintillation counter (TOF) for particle identification,  
an electromagnetic calorimeter (ECL) for photon detection, photon energy measurement, and electron identification, surrounded by a $1.5 \, {\rm T}$ superconducting solenoid, 
and resistive plate chambers installed in the flux-return yoke to detect $\KL$ and $\mu$ (KLM).
The ECL, which is crucial for this analysis, 
consists of 8736 CsI(Tl) crystals with a nearly projective geometry covering the polar angle range of
$12\degree<\theta<157\degree$. 

 To mitigate possible biases in the analysis, we establish 
 the event selection and the search method using simulation, and we validate these with control modes and off-resonance data before the experimental data are examined. 
The signal $\BtoKa$ processes are simulated using the EvtGen generator~\cite{Lange:2001uf}.
 We generate 60 samples for $\kaon$ modes, with $\ap$ masses ranging from $0.16\gev$ to $4.50\gev$, and 
 56 samples for $\Kstar$ modes, with $\ap$ masses ranging from $0.16\gev$ to $4.20\gev$. 
 The long-lived $\ap$ sample is simulated 
 with the ALP lifetime $c\tau$ varying from $10\mm$ to $500\mm$, where $\tau$ is the proper lifetime.
In addition, we simulate a sample with  uniformly distributed diphoton invariant mass in the entire mass range from $0.01\gev$ to $4.78 \gev$ for $\kaon$ modes, and $0.01\gev$ to $4.28 \gev$ for $\Kstar$ modes with $10\mev$ intervals.

Several processes contribute as background, 
including $\epem\ra\FourS\ra\BBbar$ with $B$-meson decaying to SM particles that can mimic the signal, 
as well as $\epem\ra\qqbar\ (q=u, d, s, c)$ continuum processes.
We simulate the background processes $\FourS\ra\BBbar$ and $\epem\ra\qqbar$ ($q=u,d,s,c$) using the EvtGen~\cite{Lange:2001uf} and PYTHIA~\cite{Sjostrand:2014zea} generators.
 Final-state radiation is simulated using PHOTOS~\cite{Barberio:1990ms}.
 The detector response is simulated with GEANT3~\cite{Brun:1119728}.
 Both experimental data and simulated events are converted to the Belle~II format using B2BII~\cite{Gelb:2018agf}
 and then analysed using the Belle~II analysis software framework~\cite{Kuhr:2018lps,basf2-zenodo}.

\section{Selection of signal events}

The charged particles in $\BtoKa$ decays, which originate from one of the four kaon modes $\Kp$, $\Kstarz$, $\Kstarp$, and $\KS$, 
are required to have a point of closest approach to the interaction point of less than $4\cm$ in the $z$-direction and less than $3\cm$ in the radial direction.
The charged kaons ($\Kp$) are identified using a likelihood ratio ${\cal P}$, which compares two particle hypothesis $i$ and $j$, ${\cal P}(i:j) = \mathcal{L}_i/(\mathcal{L}_i + \mathcal{L}_j)$. 
The likelihoods $\mathcal{L}$ are calculated based on the Cherenkov photon yield in the ACC, the energy-loss measurements in the CDC, and the time-of-flight information from the TOF.  
To select $\Kp$, we require ${\cal P}(\kaon:\pi) > 0.6$ and ${\cal P}(\kaon:\proton) > 0.4$.
The $\Kstar$ candidates are reconstructed from the decay modes $\Kstarz \ra \Kp \pim$ and $\Kstarp \ra \KS \pip$ within the invariant mass range of 0.8\gev to 1.0\gev.
To select charged pions $\pi^\pm$ we require ${\cal P}(\pi:\kaon) > 0.4$ and ${\cal P}(\pi:\proton) > 0.7$, with the likelihood ratio definitions being the same as for the charged kaon. 
The identification efficiencies for pions and kaons are 96.9\% and 85.3\%, respectively, with misidentification rates for pions as kaons and vice versa of 10.3\% and 1.3\%, respectively, while the misidentification rates for protons are negligible. 
The $\KS$ candidates are reconstructed from the $\KS\ra\pip\pim$ decay mode. 
Pairs of oppositely charged pions are  combined, and an artificial neural network~\cite{Belle:2022pwd} is then used to identify the $\KS$ candidates. The neural network is trained with 13 variables.
The dipion invariant mass is required to be within a range of $20\mev$ around the nominal mass of the $\KS$~\cite{ParticleDataGroup:2022pth}.
This range corresponds to approximately $\pm 5\sigma$ window around $\KS$ mass, 
where $\sigma$ is the $\pip\pim$ invariant mass resolution.
Detailed information is provided in Ref.~\cite{Nakano:2015vvh}.
Particles with high lepton probability are excluded from our analysis. 
For the electron identification, we primarily use the information from the ECL along with other subdetectors. 
The electron likelihood ratio is defined as ${\cal P}(e) = \mathcal{L}_e / (\mathcal{L}_e +\mathcal{L}_{\overline{e}})$, 
where $\mathcal{L}_e$ is the electron likelihood and $\mathcal{L}_{\overline{e}}$ is the non-electron likelihood,
a product of likelihoods from the ACC, CDC, and ECL.
For muon identification, the likelihood is calculated using the information from the KLM. The likelihood ratio is defined as 
${\cal P}(\mu) = \mathcal{L}_\mu/(\mathcal{L}_\mu +\mathcal{L}_\pi + \mathcal{L}_\kaon)$, 
where the value $\mathcal{L}_\mu$ is determined based on whether the charged particle has an associated KLM signature.
We require ${\cal P}(e) < 0.9$ and ${\cal P}(\mu) < 0.9$ to veto leptons.


The ALP candidates are reconstructed from the $\ap\ra\gaga$ decay. 
Photons are identified from ECL energy deposits that are not associated with reconstructed charged particles. 
To suppress the contribution from photons that originate from beam background, 
each photon candidate is required to  have a minimum energy that depends on the ECL region: $E_\g > 50\mev$ in the barrel ($32.2 \degree < \theta < 128.7\degree$),
 $E_\g> 100\mev$ for the forward endcap ($12.0\degree < \theta < 31.4\degree$),
 and $E_\g > 150\mev$ for the backward endcap ($131.5\degree < \theta < 157.1\degree$),
where $\theta$ is the polar angle of the photon candidate in the laboratory frame. 

The signal $B$ candidates are then formed by combining an ALP candidate with a $\kaon^{(*)}$ candidate. 
We select $B$ candidates using two kinematic variables: 
the beam-constrained mass $\Mbc = \sqrt{E_{\rm{beam}}^2 - p_B^2}$
and the energy difference $\dE = E_B-E_{\rm{beam}}$, 
where $p_B$ and $E_B$ are the momentum and energy of the $\B$ candidate in the centre-of-mass (c.m.)\ frame, and $E_{\rm{beam}}$ is the beam energy in the c.m.\ frame.
We require $\Mbc > 5.27\gev$ and $-0.2\gev < \dE < 0.1\gev$.
A kinematic fit is applied to the selected $B$ candidates, 
constraining the $B$ decay position to the interaction point, 
and the invariant mass of the $K^{(*)}\gamma \gamma$ to the nominal $B$ mass.
After $\B$ meson reconstruction, about 25\% of events have multiple signal candidates.
For such events, we select the candidate with the smallest $|\dE|$.
According to Monte Carlo simulations, approximately 85\% of the candidates are correctly reconstructed after the candidate selection.

A series of fast boosted decision tree (BDT) classifiers~\cite{Keck:2017gsv} are used after event selection to suppress the background processes. 
To train the BDT classifiers, ALP simulations with uniformly distributed diphoton invariant mass, $\Mgg$, are used.
This ensures that the classifiers do not learn or exploit the signal peak position in their background rejection strategy, and consequently avoids bias in the  search for ALPs at a specific mass hypothesis.
Since the distributions of input variables differ significantly between the low and high mass regions, 
the data above and below $1\gev$ are treated separately to enhance classification power.
 The training of all classifiers is based on simulated samples that are statistically independent of those used to develop and validate the fitting strategy. 
 This separation ensures an unbiased optimization of the background suppression procedure before its application to the unblinded data.

 We apply two continuum suppression ($CS$) classifiers to suppress the dominant background process, $\epem\to\qqbar$.
The first classifier, $CS1$, uses 10 variables.
These are the ratio of zeroth to second Fox-Wolfram moments ($R_2$)~\cite{Fox:1978vu}, 
several modified Fox-Wolfram moments~\cite{Belle:2003fgr,Belle:2001ccu}, 
the cosine of the angle between the signal $B$-thrust and the rest-of-event (ROE) thrust axes, 
the cosine of the angle between the signal $B$-thrust and the $z$ axes,
and the magnitude of the ROE thrust. 
The thrust axis is defined as the axis that maximizes the sum of the projected momenta of all particles~\cite{BaBar:2014omp}. 
The signal $B$ thrust axis is calculated using particles in the signal candidate, 
while the ROE thrust axis is determined from the charged particles and photons not used in the signal candidate reconstruction.
The second classifier, $CS2$, 
is trained using events with $CS1 > 0.1$ and utilizes the event sphericity and aplanarity (which is a linear combination of the sphericity eigenvalue and 3/2 of the third sphericity eigenvalue\cite{Bjorken:1969wi}),
the sum of the absolute value of the momenta of the particles moving along or against the thrust axis,
harmonic moments (coefficients of the spherical harmonic event expansion around the thrust axis), 
the energy asymmetry between the two photons, 
modified Fox-Wolfram moments, 
and number of ALP candidates per event ($N_{\rm cand}$).  
We find, based on our simulation study, that $N_{\rm cand}$ is larger for signal events than for background ones. 
The most discriminating variables in both classifiers are the cosine of the angle between the signal $B$-thrust and ROE-thrust axes for the low ALP mass hypothesis and the reduced Fox-Wolfram $R2$ for the high ALP mass hypothesis. 
 
A major background arises from $\piz$ mesons produced in $\Upsilon(4S)\to B \bar B$ or $e^+ e^- \to q \bar q$ processes, decaying into $\gamma \gamma $.
To suppress these, 
we first calculate identification variables by combining one photon, $\gamma_a$, from the reconstructed ALP candidate with any other photon, $\gamma_b$, in the ROE. 
For each pair of photons $(\gamma_a, \gamma_b)$, we perform three binary classification tests, each based on different assumptions about their source: 

\begin{itemize}
    \item[(1)] We compare a pair in which $\gamma_a$ is from a true ALP signal and $\gamma_b$ is from the ROE in a signal event, against a pair from the same $\piz$ decay in a background event;
    \item[(2)] We compare a pair in which $\gamma_a$ is from a true ALP signal and $\gamma_b$ is from the ROE in a signal event, against a pair where $\gamma_a$ and $\gamma_b$ are from different processes in a background event;
    \item[(3)] We compare a pair in which both $\gamma_a$ and $\gamma_b$ come from different processes in a background event, against  a pair  where both photons come from the same particle decay, such as a $\piz$, also in a background event. 
\end{itemize} 
The tests are carried out with BDTs trained with multiple variables from Monte Carlo simulated events 
such as the ratio of energies $E_9$ and $E_{25}$ in the inner $3\times3$ and $5\times5$ crystals around the central crystal ($E_9/E_{25}$),
the energy of the most energetic crystal in the ECL cluster,
the sum of weights of all crystals in an ECL cluster,
the photon energy, as well as kinetic properties of the diphoton system such as its mass, energy, opening angle, transverse momentum, and energy asymmetry. 
Among the various possible partner photons $\gamma_b$, the one that yields the highest $\piz$-like score in each test is selected.
These three individual scores are then combined into a single score, which is used to calculate the probabilities, $P_{\piz}(\g_1)$ and $P_{\piz}(\g_2)$, for each signal candidate photon, $\gamma_1$ and $\gamma_2$, to originate from a $\piz$.

For each ALP $\ap$ mass range and kaon mode, selection criteria on the four BDT classifier scores, 
$CS1$, $CS2$, $P_{\piz}(\g_1)$ and $P_{\piz}(\g_2)$,  
are optimized using the Punzi figure of merit (PFM)~\cite{Punzi:2003bu}.
 A four-dimensional grid search with a step size of $\mathcal{O}(10^{-3})$ is performed,
 selecting the points with the highest PFM value as the minimum acceptance values of the BDT score.
 The BDT selections have average signal efficiencies of 66.7\%, 87.5\%, 96.1\%, 99.0\%,
 with average background events rejection rates of 95.4\%, 31.1\%, 27.8\% and 11.1\%, respectively.    

An additional background contribution comes from the process $\B\ra\X_s\g$, where $X_s$ is any hadronic state that contains an $s$ quark.
To suppress these events, we employ a BDT classifier, ${\rm BDT}_{X_s \gamma}$, trained with six variables, separately for the low and high ALP mass regions.
These variables include the $E_9/E_{25}$ ratio of each photon,
the energy of the most energetic crystal in the ECL cluster, and two helicity angles. 
The first angle is between the photon momentum and the direction opposite to the $\B$ momentum in the $\ap$ rest frame, 
and second angle is between the $K^{(*)}$ momentum and the direction opposite to the c.m.\ system in the $\B$ rest frame.
The most powerful discriminating variables for all mass hypotheses are the first helicity angle and the energy of the most energetic photon in the c.m.\ frame.
The ${\rm BDT}_{X_s \gamma}$ score is not included in the global BDT selection optimization described above. 
Since this classifier targets a particular specific background process, global optimization may not be appropriate. 
Instead, we evaluate several threshold values ${\rm BDT}_{X_s \gamma}$ and select the one that yields the best PFM for $X_s\g$ suppression.
The BDT for $X_s\g$ suppression achieves an average signal efficiency of 76.6\%, 
with a background rejection rate of 53.0\%.
Distributions of the BDT classifier scores  can be found in Appendix~\ref{sec:B}.

The signal-selection efficiency varies across different kaon modes, and within each $K^{(*)}$ mode, 
the efficiency depends on the ALP mass, which strongly correlates with the $K^{(*)}$ momenta. 

After all selection requirements, the signal selection efficiency varies across different kaon modes.
The $\Kp$ mode shows the highest efficiency, but it also depends on the ALP mass. In the low-mass region, the  efficiency varies between 8\% and 10\%, while in the high-mass region it shows a  stronger variation, reaching a minimum of 7.5\%  and peaking at about 16\% near an ALP mass of 3 GeV. 
The $\KS$ mode follows with slightly lower efficiency between 6\% and 11\%. 
The $\Kstarz$ mode shows further reduced efficiency between 3\% and 8\%.
The $\Kstarp$ mode has the lowest efficiency, around 2\% across the entire mass range.

\begin{figure}[t]
\subfloat[\label{fig:Mgg_data-a}]{\includegraphics[width=1.0\linewidth]{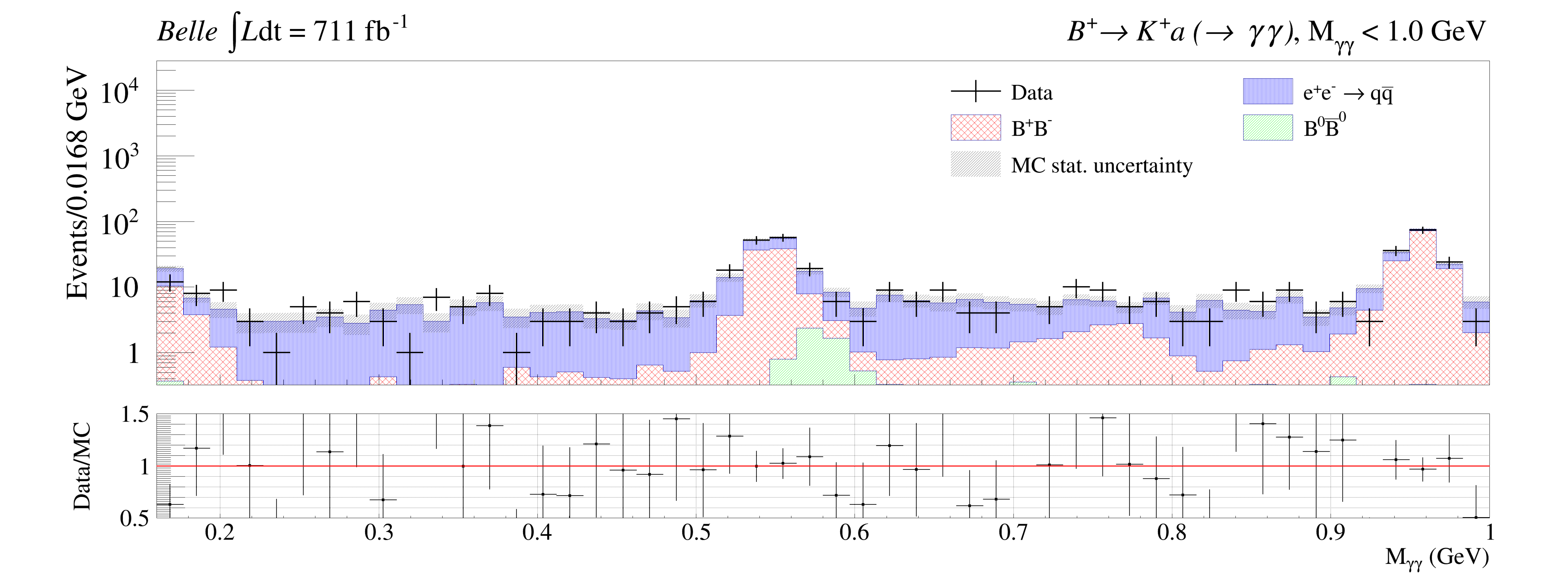}}\hfill
\subfloat[\label{fig:Mgg_data-b}]{\includegraphics[width=1.0\linewidth]{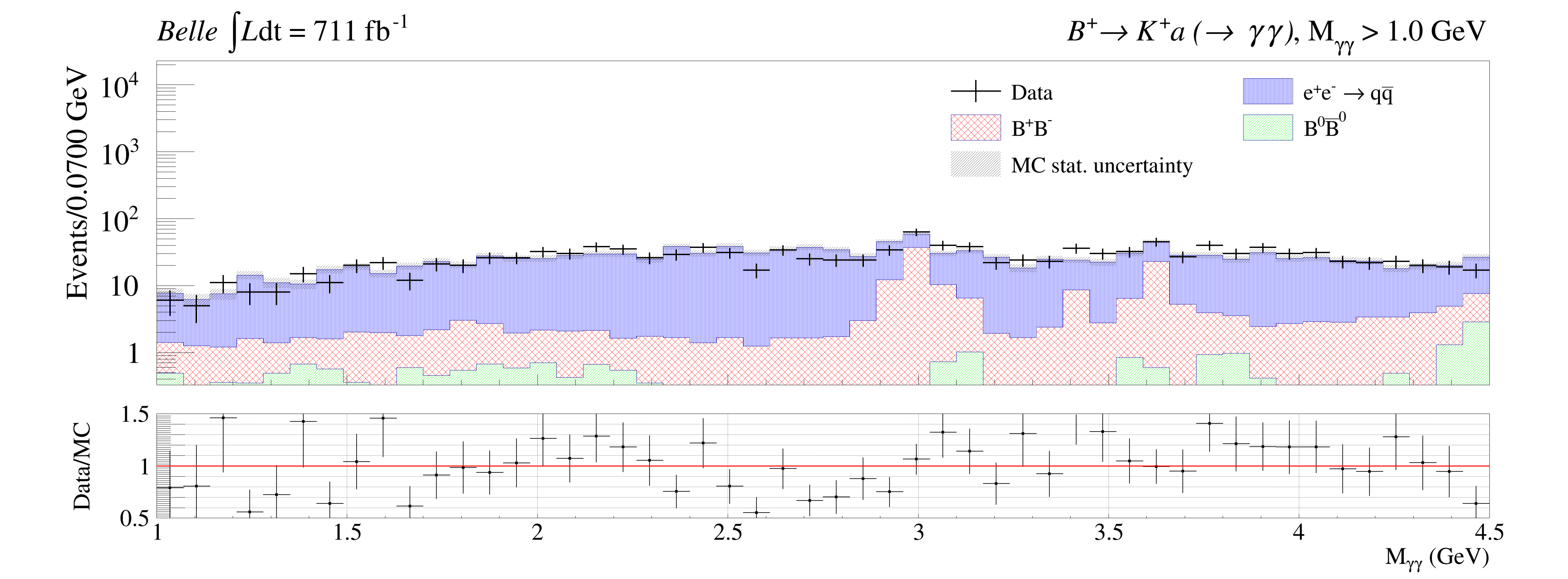}} 
\caption{\label{fig:Mgg} 
Diphoton invariant mass distribution of ALP candidates in $\Bp\ra\Kp\ap$ decay,
overlaid with simulated background contributions from $\epem\to\qqbar$ (blue vertical hatched), 
$\epem \to \FourS \to \BpBm$ (red cross-hatched),
and $\epem \to \FourS \to \BzBzb$ (green diagonal hatched) 
normalized to the experimental data luminosity, with all weights applied.}
\end{figure}

\section{Simulation calibration} 
We use the off-resonance data to evaluate corrections to the simulation by comparing experimental data with simulated distributions.
 The off-resonance data are collected 60\mev below the $\FourS$ resonance energy, 
 where $\epem$ collisions produce all processes except for $\BBbar$ pair production.
There are two types of discrepancy found between the experimental data and the simulation:
the number of background events, and the shape of the event kinematic and topology variables. 
For the former, we derive the ratios between the off-resonance data and simulation for each ALP mass range as weights, 
which are then applied to both the off-resonance and on-resonance continuum simulations.
For the latter, we adopt a data-driven method~\cite{Martschei:2012pr}.
A BDT classifier is trained to distinguish the differences between the off-resonance data and simulation, 
using the same variables applied for continuum suppression.
Given the BDT output $p(i) $ with $0<p(i)<1$ for each simulation event $i$, 
we calculate the weight $w(i) = p(i)/(1-p(i))$, which is then applied as a ``continuum shape correction'' factor.  
This procedure gives greater importance to simulated events that most closely resemble the real data, 
improving the agreement between the weighted simulated  and  experimental distributions. 
Reweighting based on the off-resonance data enhances the simulated two-photon invariant mass distribution, 
which then matches the experimental distribution within statistical uncertainty.
The resulting  on-resonance $\Mgg$ distributions for low and high ALP mass regions in the $\Bp\ra\Kp\ap$ decay are shown in Fig.~\ref{fig:Mgg}.

\section{Signal extraction and validations}

\begin{figure}[t]
\subfloat[\label{fig:Ext0}]{\includegraphics[width=.5\linewidth]{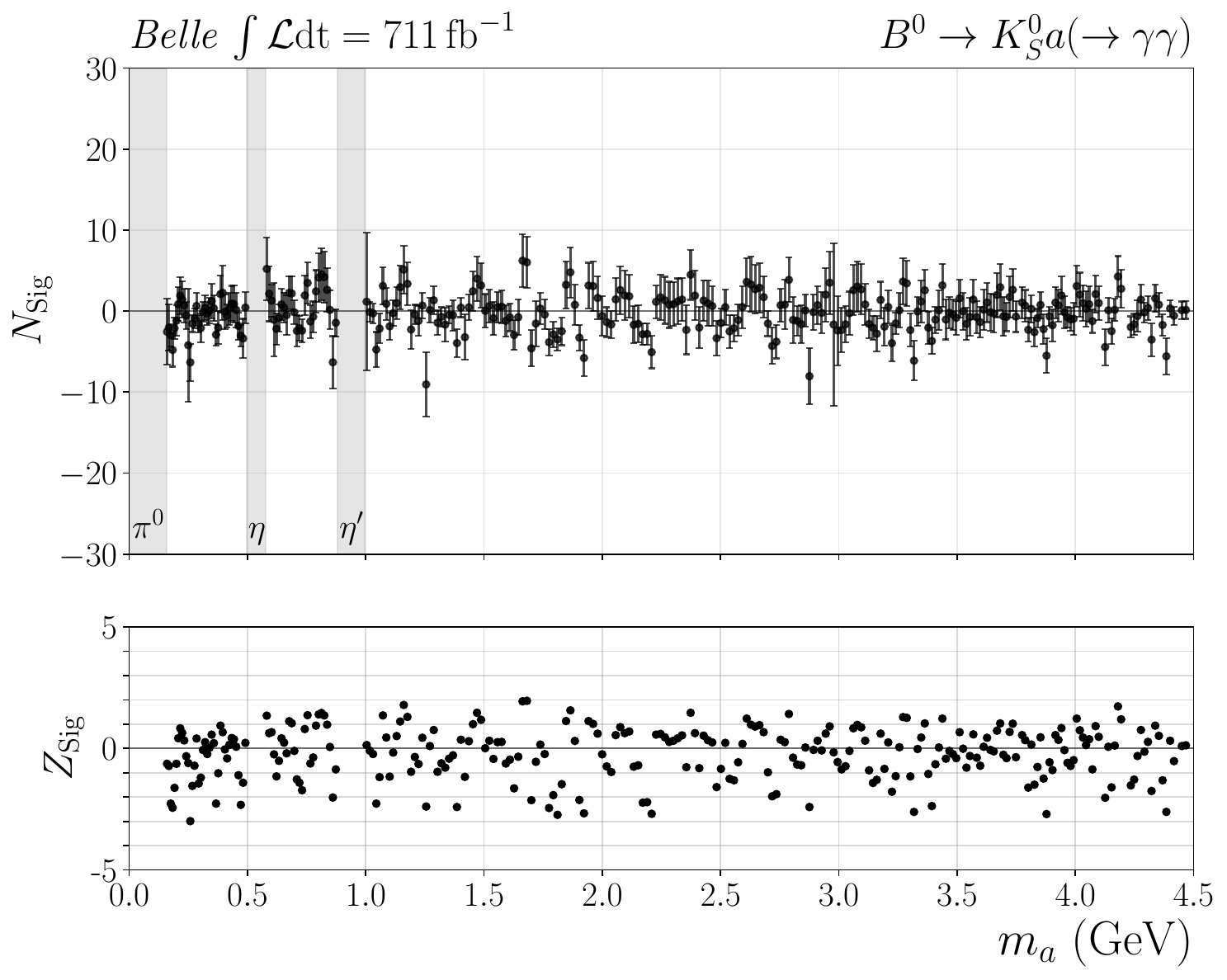}}
\subfloat[\label{fig:Ext1}]{\includegraphics[width=.5\linewidth]{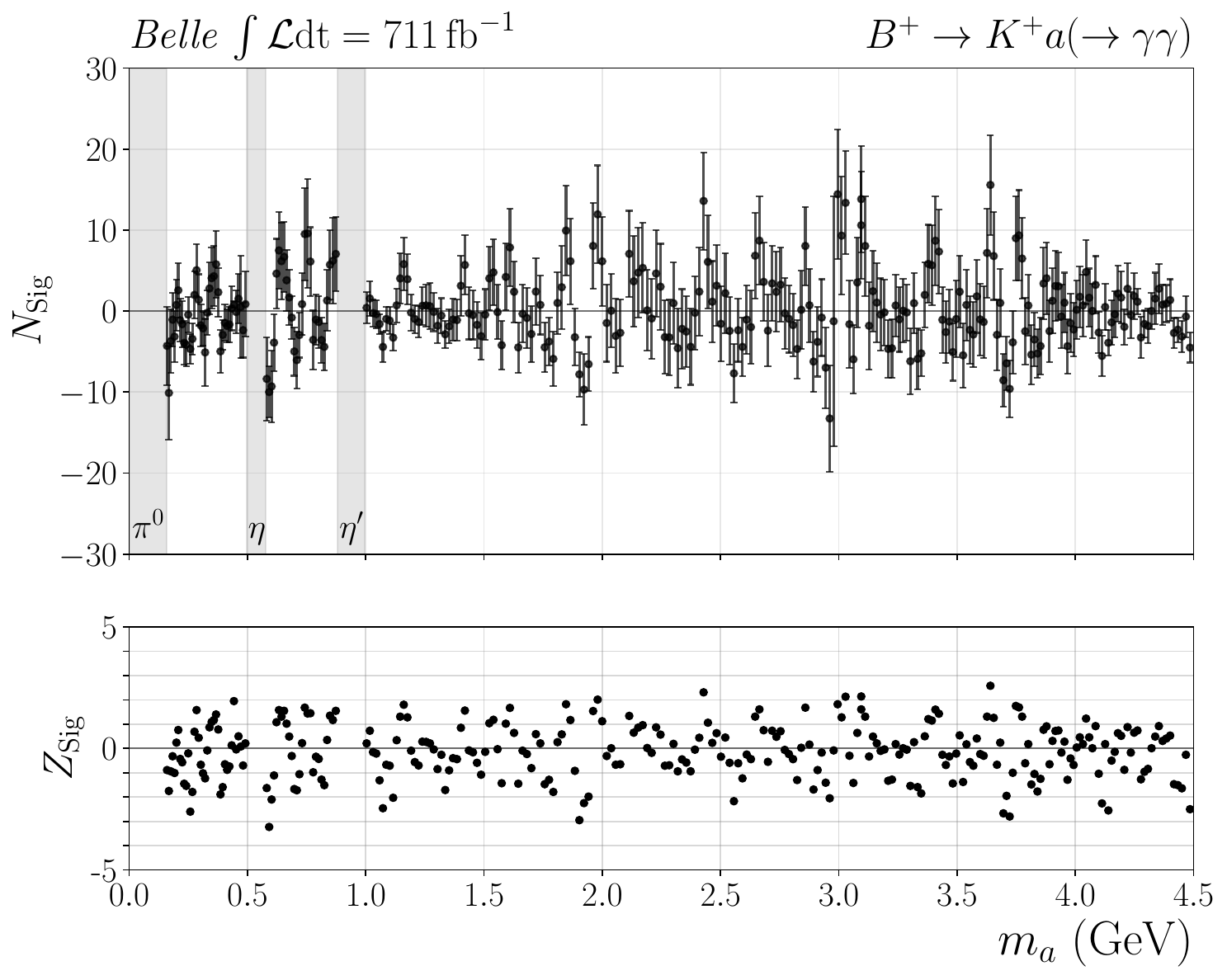}}\hfill
\subfloat[\label{fig:Ext2}]{\includegraphics[width=.5\linewidth]{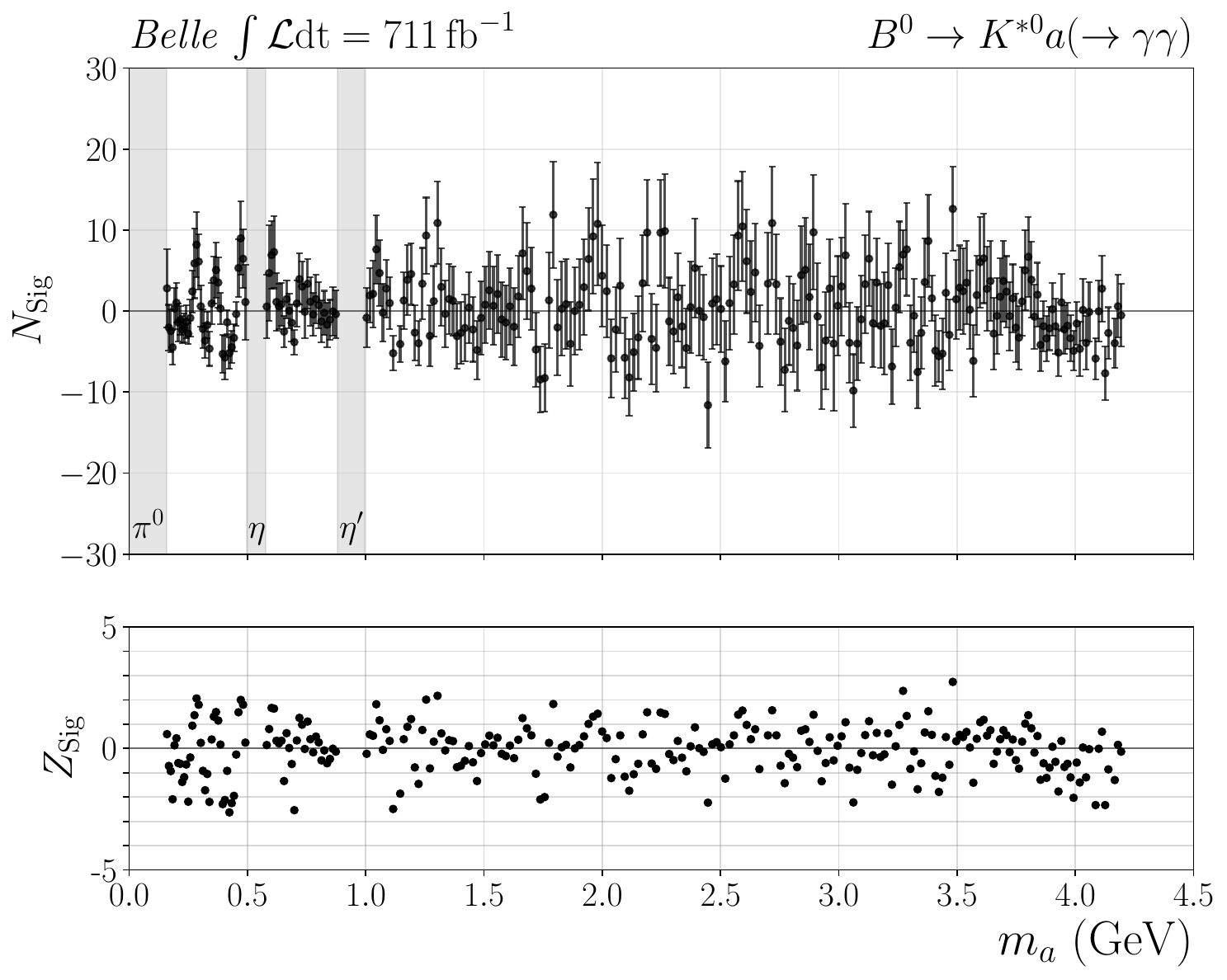}}
\subfloat[\label{fig:Ext3}]{\includegraphics[width=.5\linewidth]{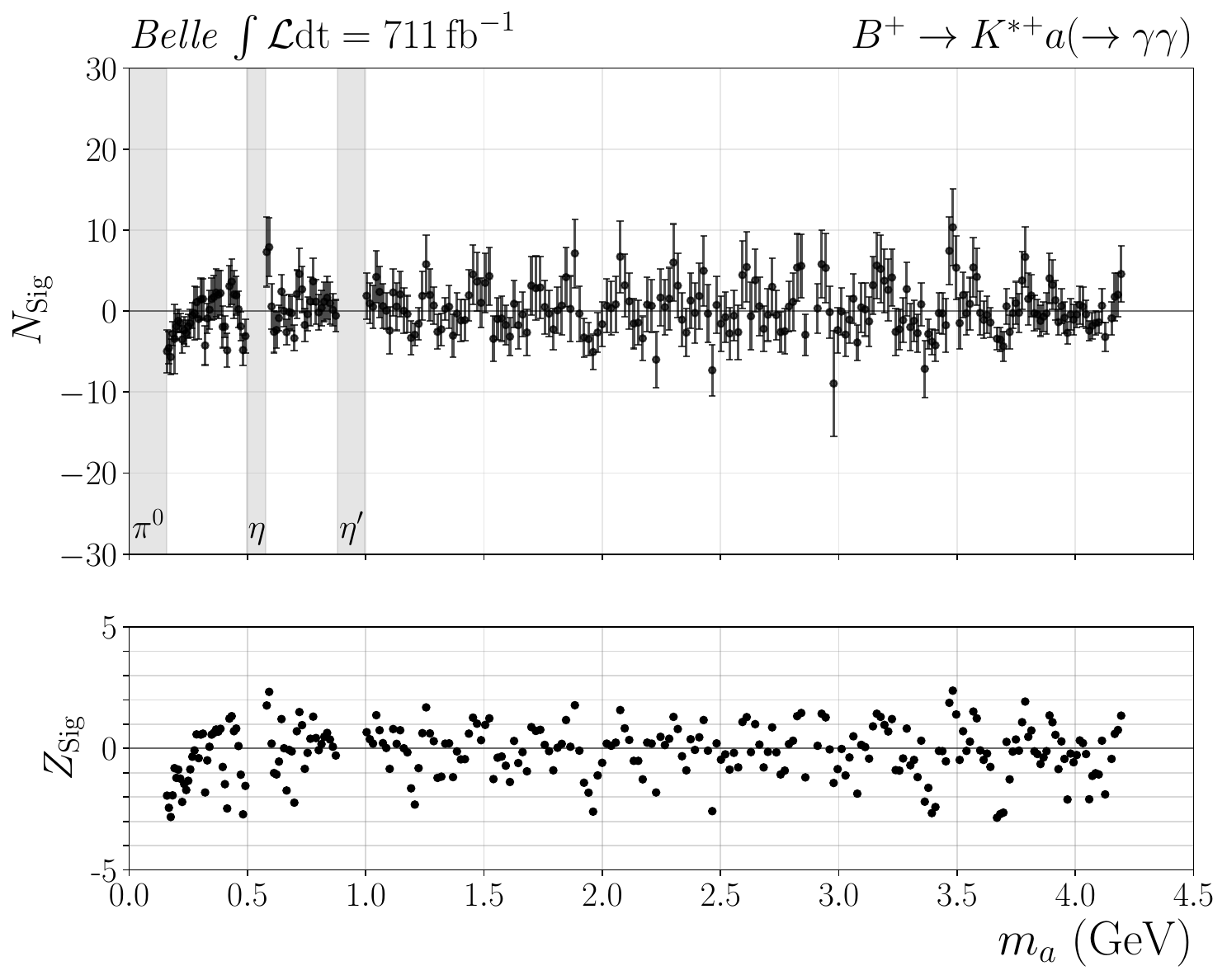}}\hfill
\caption{\label{fig:Ext} 
Extracted signal yield ($N_{\rm Sig}$) (top)
and the significance level ($Z_{\rm Sig}$) (bottom) for four kaon modes.
The grey bands are the excluded regions corresponding to the $\piz,\eta$ and $\etp$ mass regions.}
\end{figure}

 The signal yield $N_{\rm Sig}$ and its standard deviation $\sigma_{\rm Sig}$ are obtained from an
 unbinned maximum likelihood fit to the $\Mgg$ distribution. 
 We use a  double-sided Crystal Ball function~\cite{Gaiser:Phd} to model the signal $M_{\gamma\gamma}$ distribution.
 The non peaking background in the $\Mgg$ distribution  is parametrized using a second-order polynomial.
The peaking background components from $h\to \gamma \gamma$ ($h = \piz$, $\eta$, or $\etap$) are modelled with  double-sided Crystal Ball functions.

We perform a mass scan with a step size equal to the high-side mass resolution parameter from the Crystal Ball signal, $\sigma_{\gaga}^R$.
The latter varies with the ALP mass, ranging from $7.8\mev$ at $\ma = 0.160 \gev$ to
$19.4 \mev$ at $\ma = 1.9 \gev$, and decreasing to $17.9 \mev$ at $\ma = 4.5\gev$.  
The improved mass resolution in the highest $m_a$ region  results from kinematic fitting. 
Each fit range extends over an $\Mgg$ interval with a width
of $9\times(\sigma_{\gaga}^R+\sigma_{\gaga}^L)$, where  $\sigma_{\gaga}^L$ is the low-side Crystal Ball resolution parameter.
The signal peak shape parameters depend on the ALP mass, and are derived from the corresponding signal samples. 
The shape parameters and position of the peaking background are fixed based on values obtained from the simulation. 
The combinatorial background parameters  are floating in the fit, 
along with the signal and peaking background normalization, other than the $\etc$. 

Due to the peaking background from $\piz$, $\eta$ and $\etp$ decays, masses below 0.160\gev, in the ranges 0.497\gev--0.578\gev and 0.938--0.997\gev, are excluded from the ALP mass scan.
The latter two correspond to $\pm3\sigma$ mass resolution windows centred on the $\eta$ and $\etp$, respectively. 
Note that despite these exclusions, the tails of these peaking background components extend into the fitting regions.
The ALP mass ranges close to the kinematic limit,  
from $4.50\gev$ to $4.78\gev$ for $\kaon$ and from $4.20\gev$ to $4.38\gev$ for $\Kstar$, are excluded from our analysis due to low signal efficiency and insufficient background population for reliable signal extraction. 
The $\etc$ mass region is included in the ALP mass scan, as the branching fraction of $\B \to \kaon^{(*)}\etc(\to \gaga)$ is sufficiently small that its contribution can be adequately controlled.
The normalization factor for the $\FourS\ra\BBbar$ background component involving $\etc$ is constrained to the world-average value within the SM uncertainty, 
while the normalization of the $\etc$ component  in the $\epem\ra\qqbar$ background is treated as a free parameter in the fit.

To validate the signal extraction method, we measure the branching fraction of $\B\ra\kaon h (\ra\gaga)$ using the  fitting procedure described above.
The branching fractions of the $\Bp\to\Kp\eta$, $\Bp\to\Kp\etp$ and $\Bp\to\Kp\etc$ decays are measured to be $(0.95\pm0.27)\times 10^{-6}$, $(1.79\pm0.45)\times10^{-6}$ and $(2.37\pm0.91)\times 10^{-7}$, respectively, 
consistent with both the previous Belle results~\cite{Belle:2006mzv} and world average values  excluding the Belle result~\cite{ParticleDataGroup:2022pth}.
The parametrisation of signal shapes are also validated with these control modes.  
In all control modes, the shape parameters measured from data and simulated samples are compatible.

In addition, a toy Monte Carlo (ToyMC) study~\cite{Verkerke:2003ir} --- a simplified, fast simulation in which observables are sampled from probability density functions (p.d.f.) --- is carried out to evaluate the fitting bias and signal sensitivity.
Ten thousand pseudo-datasets are generated for each ALP mass hypothesis across the four kaon modes  using the fitted p.d.f. and the Poisson distribution of signal and background yield obtained from the fit.
The result of fits on pseudo-data from the ToyMC has a small negative bias of $4.2\%\times\sigma_{\rm Sig}$ on average, 
where $\sigma_{\rm Sig}$ is the width of the signal yield distributions obtained with the ToyMC study.  
This is applied as a correction factor to the signal yields in data.

\begin{figure}[t]
 \includegraphics[width=1.0\linewidth]{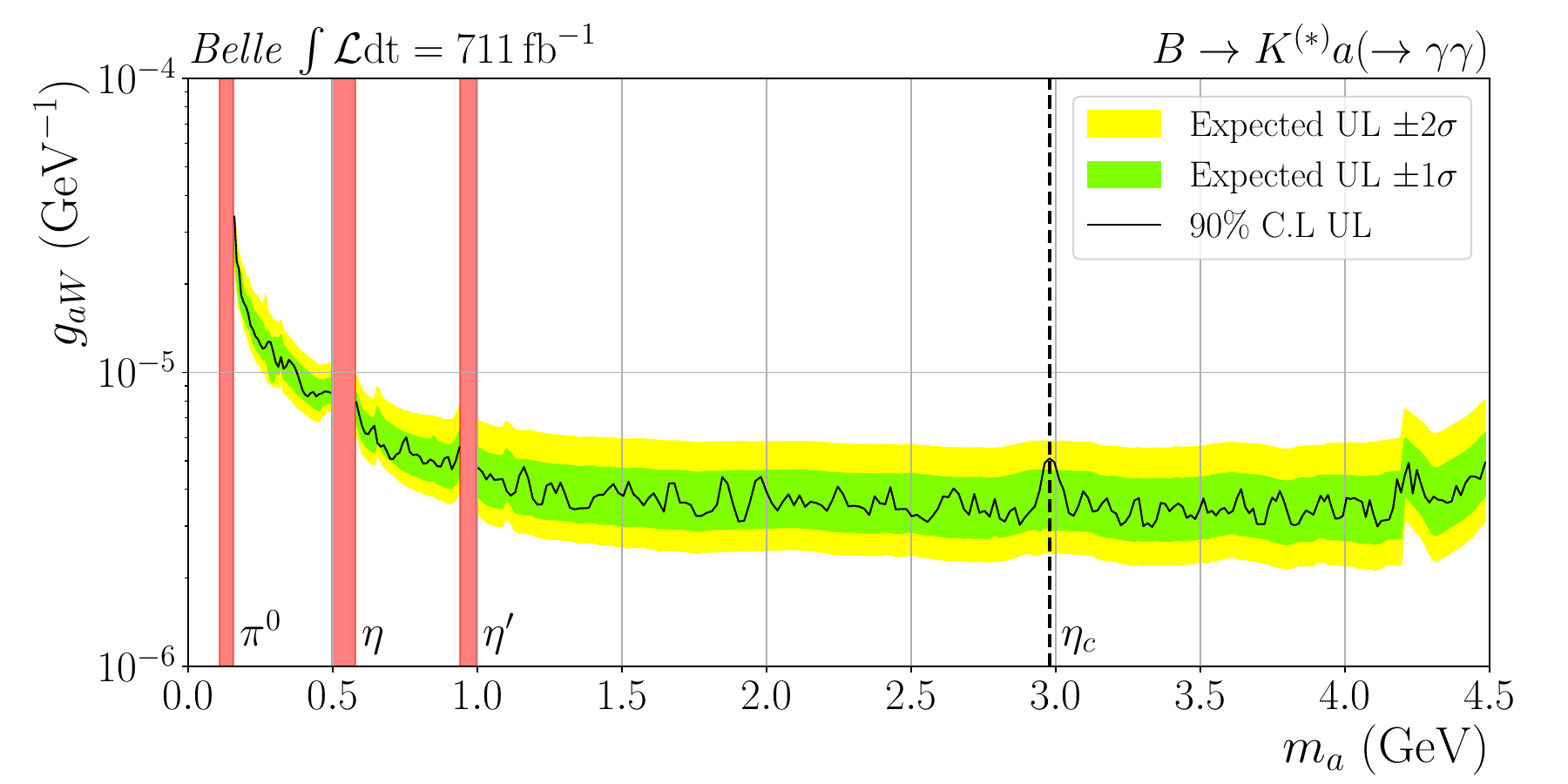}
 \caption{\label{fig:gaWUL} 
90\% CL upper limits on the coupling $\gaW$ as a function of the ALP mass obtained with the CLs method with simultaneous fit to the four kaon modes. 
 The green and yellow bands are the $\pm 1$ and $\pm 2$ standard deviation ranges, respectively, 
 for the expected upper limits in the background only model. 
 The red bands are the excluded $\piz,\eta$ and $\etp$ mass regions.
 The vertical dashed line indicates the nominal $\etc$ mass.
 Systematic uncertainties are included in the figure.}
 \end{figure}
 
For all ALP mass hypotheses and kaon modes, the decay yield of $\BtoKa$ is measured under the assumption that all signal events originate from the prompt decay of the ALP.
However, the ALP can be long-lived, and the displaced vertex  reduces the signal efficiency.
The primary reason for the reduced signal efficiency is the assumption that the photons 
originate from the IP. The calculated opening angle between the two photons, which enters the ALP mass calculation, is systematically smaller than the true value, which produces a low-side tail on the reconstructed mass distribution.
The fit using Crystal Ball parameters from prompt decays systematically underestimates the number of signal events for displaced vertices, effectively reducing the efficiency.
To quantify this effect, signal processes with lifetimes $c\tau$ of $10\mm, 50\mm, 100\mm, 200\mm, 300\mm, 400\mm$ and $500\mm$ are generated.
The decrease in the signal efficiency is modelled as 
\begin{equation}
\label{eq:LLPeff}
\frac{\varepsilon_{\rm Sig}(c\tau)}{\varepsilon_{\rm Sig}(0)} = re^{a_1c\tau} + (1-r)e^{a_2c\tau} ,
\end{equation}
where $\varepsilon_{\rm Sig}(0)$ and $\varepsilon_{\rm Sig}(c\tau)$ are the reconstruction efficiencies of prompt and long-lived ALPs,  $a_1$, $a_2$ and $r$ are floating parameters obtained from fits to the long-lived ALP simulation results for each ALP mass hypothesis. 
The resulting functions are incorporated into the upper limit calculation.
Consequently, ALP signals with relatively low mass have a longer lifetime and lower signal efficiency, leading to less stringent limits on $\gaW$. 
 
The significance is evaluated as $Z_{\rm Sig} = \sqrt{2(L_{\rm s+b} - L_{\rm b})}$, where $L_{\rm s+b}$ and $L_{\rm b}$ are the negative log-likelihoods of the fits with and without signal hypothesis, respectively. 
The largest observed positive local significance  is $2.74\sigma$ at $\ma = 3.482 \gev$ in the $\Kstarz$ mode. 
This local significance corresponds to $1.89\sigma$ global significance, after including the look-elsewhere effect~\cite{Gross:2010qma}.
As shown in Fig.~\ref{fig:Ext}, no significant excess over background is observed and we set 90\% confidence level (CL) upper limits on the coupling $\gaW$ using the CLs method~\cite{Cowan:2010js,Read:2002hq}.
For each kaon mode and ALP mass hypothesis, we obtain the  branching fractions of $\BtoKa$ decays as
\begin{equation} 
\begin{aligned}   
\mathcal{B}(\BtoKa) &= \frac{N_{\rm Sig}}{(2\times N_{\FourS}\times f^x\times\varepsilon_{\rm Sig})}, 
\end{aligned}   
\end{equation}
where $N_{\FourS}$ is the total number of $\FourS$ mesons, 
$f^x$ is the production fraction of $\BBbar$ pairs for the neutral mode, $f^{00}$, or the charged mode, $f^{+-}$
~\cite{Banerjee:2024znd}, and $\varepsilon_{\rm Sig}$ is the signal efficiency.
For each ALP mass hypothesis, a simultaneous fit is performed on four kaon modes to obtain $\gaW$.
Figure~\ref{fig:gaWUL} shows the resulting limit on the coupling constant $\gaW$ as a function of mass.
The expected limits obtained with the background only hypothesis are also shown as green and yellow bands.
For $\ma > 4.2\gev$, only the $\Kp$ and $\KS$ modes contribute to the upper limit 
calculation, as the $\Kstar$ modes do not extend to this mass range. 
The apparent behaviour observed near $4.2 \gev$ reflects this.

\section{Systematic uncertainties}

\begin{table}[t]
\footnotesize
    \caption[Systematics]{Summary of the systematic uncertainties (\%). The ``Prompt decay signal '' term is derived from a control sample: it includes the indented items that follow, as well as additional effects (see the text for details).}
    \begin{tabularx}{1.\textwidth}{
        >{\raggedright\arraybackslash}X
        >{\centering\arraybackslash\hsize=.3\hsize}X
        >{\centering\arraybackslash\hsize=.3\hsize}X
        >{\centering\arraybackslash\hsize=.3\hsize}X
        >{\centering\arraybackslash\hsize=.3\hsize}X                                      }
        \hline                                                                            
        Source                          & $\KS$ mode    & $\Kp$ mode    & $\Kstarz$ mode& $\Kstarp$ mode\\                                \hline
        Prompt decay signal            & \multicolumn{4}{c}{$21.0$}    \\
        $~\vdash$ Continuum shape correction~\cite{Belle:2019iji}      & \multicolumn{4}{c}{$4.1$}    \\
        $~\vdash$ Photon-detection efficiency~\cite{Belle:2022apl}     & \multicolumn{4}{c}{$4.0$}    \\
        $~\vdash$ $\Kp$ identification efficiency & -             & $3.6$         & $3.6$     &     -             \\
        $~\vdash$ $f^{00}$ or $f^{+-}$~\cite{Banerjee:2024znd}    & $1.7$    & $2.1$ & $1.7$ &  $2.1$         \\
        $~\vdash$ $\KS$ reconstruction efficiency~\cite{Dash:2017heu} & $1.6$         & -             & -             &     $1.6$         \\
        $~\vdash$ $N_{\FourS}$~\cite{Belle:2013hlo}                       & \multicolumn{4}{c}{$1.4$}    \\
        $~\vdash$ Tracking efficiency~\cite{Belle:2022apl}             & $0.7$         & $0.4$         & $0.7$         &     $1.1$ \\
        Long-lived ALP efficiency      & \multicolumn{4}{c}{$7.4$}    \\

        \hline
        Total                 & \multicolumn{4}{c}{$22.3$} \\
        \hline
    \end{tabularx}
    \label{table:systTable}
 \end{table}

The systematic uncertainties that affect the extraction of $\gaW$ can be classified into two main categories: those originating from prompt decay signal  and from long-lived ALP efficiency (Table~\ref{table:systTable}). 

The uncertainty on the prompt decay signal  is derived from our most precise control sample, $\Bz\ra\Kstarz\eta$. 
The obtained branching fraction $\mathcal{B}(\Bz\ra\Kstarz\eta) \times \mathcal{B}(\eta\to\gamma\gamma) = (6.8\pm1.3)\times10^{-6}$ is in good agreement with the world-average value of $(6.3\pm0.4)\times10^{-6}$~\cite{ParticleDataGroup:2022pth}. 
We therefore assign a systematic uncertainty of 21\%, calculated as the quadratic sum of the fractional branching fraction discrepancy and the associated uncertainties, as a conservative estimate.
The indented items listed under the prompt decay signal  in Table~\ref{table:systTable} represent numerically estimable components that affect the prompt decay signal  calculation. 
Other sources not explicitly listed in the table, such as MVA selection efficiency, 
could not be individually determined and are thus incorporated into the total prompt decay signal  alongside the indented items. 
Consequently, the quadratic sum of the individual components differs from the overall prompt decay signal  value.

The systematic uncertainty of the long-lived ALP signal efficiency constraint is derived from $\KS\to\piz\piz$ decays reconstructed in $\Dstarp\to\Dz(\to\KS\pip\pim)\pip$ events by treating the $\KS$ as ALP.
Using the same method used for long-lived ALP decays (Eq.~\ref{eq:LLPeff}), 
we construct the signal efficiency function for $\KS\to\piz\piz$ and determine the expected mass resolution on $M(\piz\piz)$, 
the invariant mass of the $\piz\piz$ system, from this function and $\KS$ simulation with zero lifetime.
The $M(\piz\piz)$ mass resolution is extracted via p.d.f. fitting in both data and simulation, 
following the same approach used in ALP signal extraction. 
The fractional difference between the obtained value and the expectation is $(6.0\pm4.4)\%$, which, by adding the difference and its error in quadrature, leads to a conservative estimation of the systematic uncertainty of $7.4\%$ for the long-lived ALP efficiency calculation.
The efficiency for long-lived ALP signals decreases as the ALP mass increases, since higher masses correspond to shorter lifetimes.
This effect becomes negligibly small for masses above 2.0 GeV. 
However, as our analysis is dominated by statistical uncertainties, and since mass-dependent refinements to systematic uncertainties would not significantly impact the final results, 
we have applied the most conservative systematic uncertainty estimate across the entire mass range in our calculations.
 
\begin{figure}[t]
\includegraphics[width=1.0\linewidth]{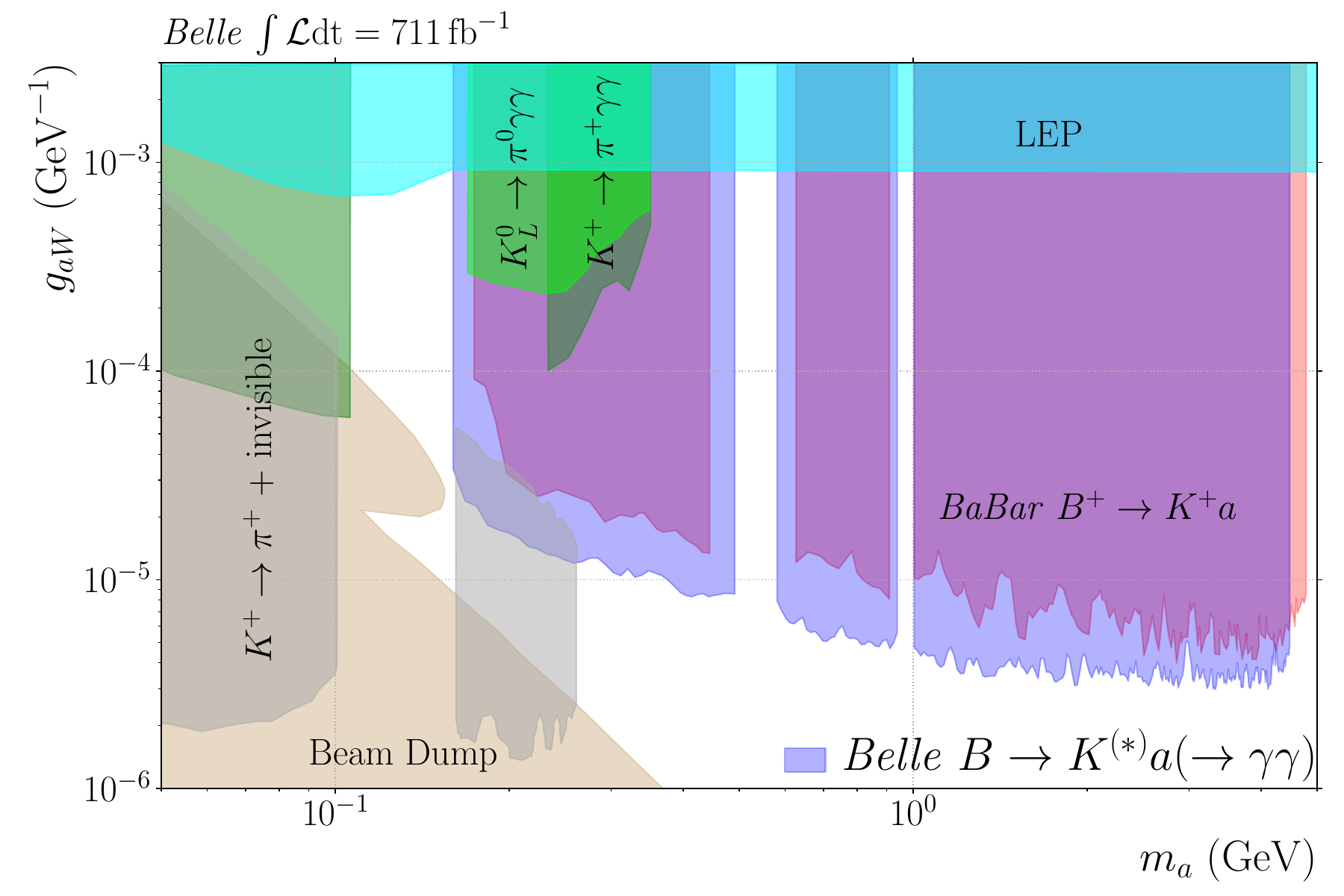} 
\caption{\label{fig:gaWExc} 
The 90\% CL upper limits on the coupling $\gaW$ 
from a simultaneous fit to the four $\B\to\kaon^{(*)}\ap$ modes as a function of the ALP mass,
compared with existing constraints~\cite{BaBar:2021ich,PhysRevLett.118.111802,NA62:2021zjw,Goudzovski:2022vbt}.}
\end{figure}

The total systematic uncertainty is obtained by summing the 
prompt decay signal and long-lived ALP efficiency uncertainties in quadrature.
The resulting systematic uncertainty is found to be 22.3\% for all kaon modes and ALP mass hypotheses. 
The resulting systematic uncertainty is included in the upper limit calculation procedure by convolving an appropriate Gaussian function with the signal likelihood.

\section{Result and conclusion}

We report a search for an axion-like particle in $\B\ra\kaon^{(*)}\ap(\ra\gaga)$ decays using a $711\invfb$ data sample collected by the Belle experiment at the KEKB $e^+ e^-$ collider at a centre-of-mass energy of 10.58\gev. 
We search for the decay of the axion-like particle into a pair of photons, $\ap\to\gaga$, and explore four kaon modes, $\KS$, $\Kp$, $\Kstarz$ and $\Kstarp$. 
We scan the two-photon invariant mass 
in the range 0.16\gev--4.50\gev for the $\kaon$ modes and 0.16\gev--4.20\gev for the $\Kstar$ modes. 
No significant signal is observed in any of the modes.  
Figure~\ref{fig:gaWExc} shows the resulting 90\% confidence level upper limits on the $\gaW$ coupling as a function of $m_a$, derived from the combination of four kaon modes. 
 The limits are $3 \times 10^{-6} \gev^{-1}$ for the ALP mass hypotheses above $2.0\gev$, increasing to $3 \times 10^{-5} \gev^{-1}$ at the lowest ALP mass.
This trend is due to the increase in the lifetime, which leads to a lower signal efficiency.
Figure~\ref{fig:gaWExc} also shows the constraints derived from the NA62 $\Kp\ra\pip+~\rm{invisible}$ search~\cite{NA62:2021zjw}. 
Based on the methodology presented in Ref.~\cite{Goudzovski:2022vbt} we reinterpret the NA62 results on a dark scalar decaying to SM
particles as limits on ALPs.
The constraints on the coupling of the axion-like particle to electroweak gauge bosons $\gaW$ are improved by a factor of two compared to the most stringent previous experimental results.\cite{BaBar:2021ich}

This work, based on data collected using the Belle detector, which was
operated until June 2010, was supported by
the Ministry of Education, Culture, Sports, Science, and
Technology (MEXT) of Japan, the Japan Society for the
Promotion of Science (JSPS), and the Tau-Lepton Physics
Research Center of Nagoya University;
the Australian Research Council including grants
DP210101900, 
DP210102831, 
DE220100462, 
LE210100098, 
LE230100085; 
Austrian Federal Ministry of Education, Science and Research (FWF) and
FWF Austrian Science Fund No.~P~31361-N36;
National Key R\&D Program of China under Contract No.~2022YFA1601903,
National Natural Science Foundation of China and research grants
No.~11575017,
No.~11761141009,
No.~11705209,
No.~11975076,
No.~12135005,
No.~12150004,
No.~12161141008,
and
No.~12175041,
and Shandong Provincial Natural Science Foundation Project ZR2022JQ02;
the Czech Science Foundation Grant No. 22-18469S;
Horizon 2020 ERC Advanced Grant No.~884719 and ERC Starting Grant No.~947006 ``InterLeptons'' (European Union);
the Carl Zeiss Foundation, the Deutsche Forschungsgemeinschaft, the
Excellence Cluster Universe, and the VolkswagenStiftung;
the Department of Atomic Energy (Project Identification No. RTI 4002), the Department of Science and Technology of India,
and the UPES (India) SEED finding programs Nos. UPES/R\&D-SEED-INFRA/17052023/01 and UPES/R\&D-SOE/20062022/06;
the Istituto Nazionale di Fisica Nucleare of Italy;
National Research Foundation (NRF) of Korea Grant
Nos.~2016R1\-D1A1B\-02012900, 2018R1\-A2B\-3003643,
2018R1\-A6A1A\-06024970, RS\-2022\-00197659,
2019R1\-I1A3A\-01058933, 2021R1\-A6A1A\-03043957,
2021R1\-F1A\-1060423, 2021R1\-F1A\-1064008, 2022R1\-A2C\-1003993;
Radiation Science Research Institute, Foreign Large-size Research Facility Application Supporting project, the Global Science Experimental Data Hub Center of the Korea Institute of Science and Technology Information and KREONET/GLORIAD;
the Polish Ministry of Science and Higher Education and
the National Science Center;
the Ministry of Science and Higher Education of the Russian Federation
and the HSE University Basic Research Program, Moscow; 
University of Tabuk research grants
S-1440-0321, S-0256-1438, and S-0280-1439 (Saudi Arabia);
the Slovenian Research Agency Grant Nos. J1-50010 and P1-0135;
Ikerbasque, Basque Foundation for Science, and the State Agency for Research
of the Spanish Ministry of Science and Innovation through Grant No. PID2022-136510NB-C33 (Spain);
the Swiss National Science Foundation;
the Ministry of Education and the National Science and Technology Council of Taiwan;
and the United States Department of Energy and the National Science Foundation.
These acknowledgements are not to be interpreted as an endorsement of any
statement made by any of our institutes, funding agencies, governments, or
their representatives.
We thank the KEKB group for the excellent operation of the
accelerator; the KEK cryogenics group for the efficient
operation of the solenoid; and the KEK computer group and the Pacific Northwest National
Laboratory (PNNL) Environmental Molecular Sciences Laboratory (EMSL)
computing group for strong computing support; and the National
Institute of Informatics, and Science Information NETwork 6 (SINET6) for
valuable network support.

\bibliographystyle{JHEP}
\bibliography{references}

\newpage
\appendix
\section{Decay width of \texorpdfstring{$\BtoKa$}{B to K a} and coupling strength \texorpdfstring{$\gaW$}{gaW}}
\label{sec:A}

The $\BtoKa$ decay width is given by
\begin{equation}\begin{aligned}
    \Gamma(\B\to\kaon\ap) &= \frac{M_\B^3}{64\pi} |g_{\ap bs}|^2 (1-\frac{M_\kaon^2}{M_\B^2})^2 f_0^2(\ma^2) \lambda_{\kaon\ap}^{1/2}, \\%
    \Gamma(\B\to\Kstar\ap) &=\frac{M_\B^3}{64\pi} |g_{\ap bs}|^2 A_0^2(\ma^2) \lambda_{\Kstar\ap}^{3/2}, \\%
    \label{eq:BF}
\end{aligned}   \end{equation}
where
\begin{equation}\begin{aligned}
    g_{\ap bs} &=
    -\frac{3\sqrt{2} G_F M^2_W g_{\ap W}}{16\pi^2} \sum_{\alpha=c,t} V_{\alpha b} V_{\alpha s}^* f(\frac{M_\alpha^2}{M_W^2}), \\%
    f(x) &\equiv \frac{x[1+x(\log x -1)]}{(1-x)^2}, \\%
    f_0(m_{\ap}^2) &= \frac{0.330}{1-\ma^2/37.46},\\%
    A_0(m_{\ap}^2) &= \frac{1.364}{1-\ma^2/27.88} - \frac{0.990}{1-\ma^2/36.78},\\%
    \lambda_{\kaon^{(*)} \ap} &= (1-\frac{(m_\ap + M_{\kaon^{(*)}})^2}{M_B^2})(1-\frac{(\ma - M_{\kaon^{(*)}})^2}{M_B^2}). \\%
    \label{eq:gaW}
\end{aligned}   \end{equation}
Here, $M_B$, $M_K$ and $M_W$ are the masses of $B$ meson, kaon and $W$ boson, respectively, while $m_{\ap}$ is the ALP mass.
The functions $f_0(q)$ and $A_0(q)$ are  the form factors from the hadronic matrix elements~\cite{Ball:2004rg,Ball:2004ye},
$G_F$ is the Fermi constant, and $V_{\alpha b}$ and $V_{\alpha s }$ with $\alpha = c, t$ are the corresponding Cabibbo-Kobayashi-Maskawa (CKM) matrix elements.

\newpage

\section{Figures}\label{sec:B}

The following includes additional figures.

\begin{figure}[ht]
\includegraphics[width=1.0\linewidth]{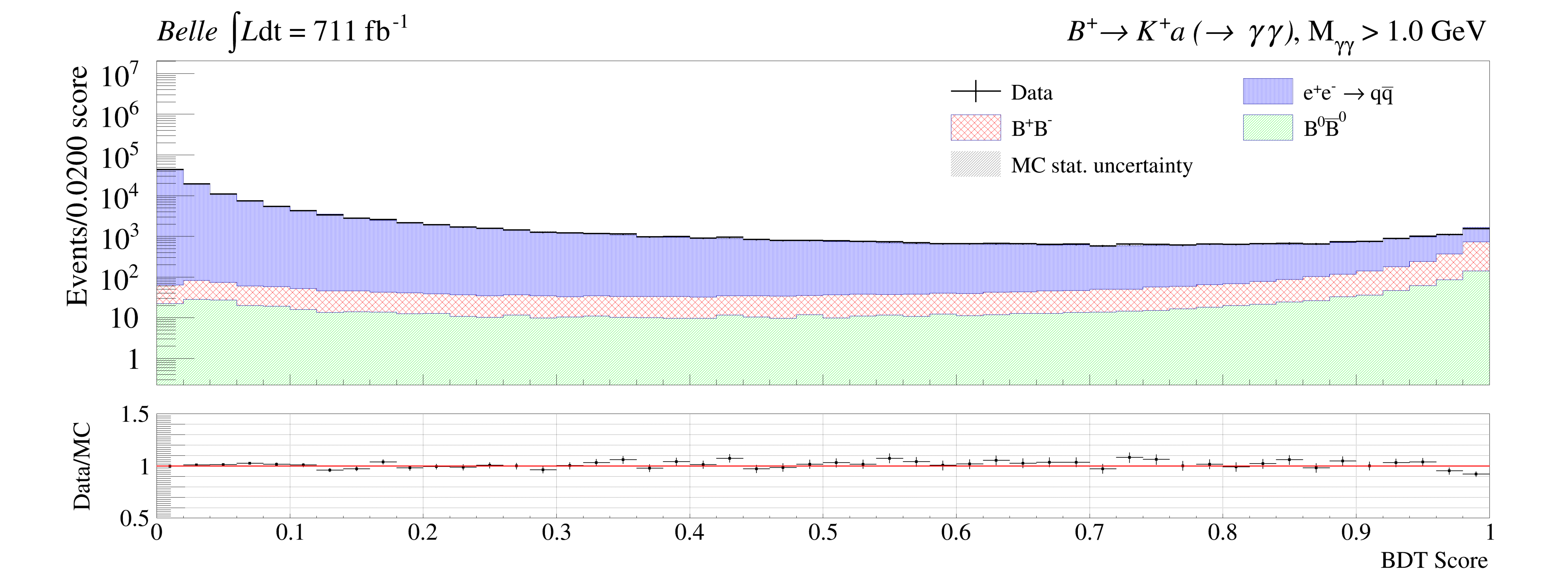}
\caption{\label{fig:BDT-a} 
Distributions of the first continuum suppression BDT classifier scores 
from experimental data (black points with error bars) 
for the $\Bp\ra\Kp\ap(\ra\gaga)$ decay with $\ma > 1.0\gevcc$ 
along with simulated background contributions from $\epem\to\qqbar$  (blue vertical hatched), 
$\epem \to \FourS \to \BpBm$ (red cross-hatched),
and $\epem \to \FourS \to \BzBzb$ (green diagonal hatched) 
normalized to the experimental data luminosity.
All correction weights are applied to the continuum events.
}
\end{figure}

\begin{figure}[ht]
\includegraphics[width=1.0\linewidth]{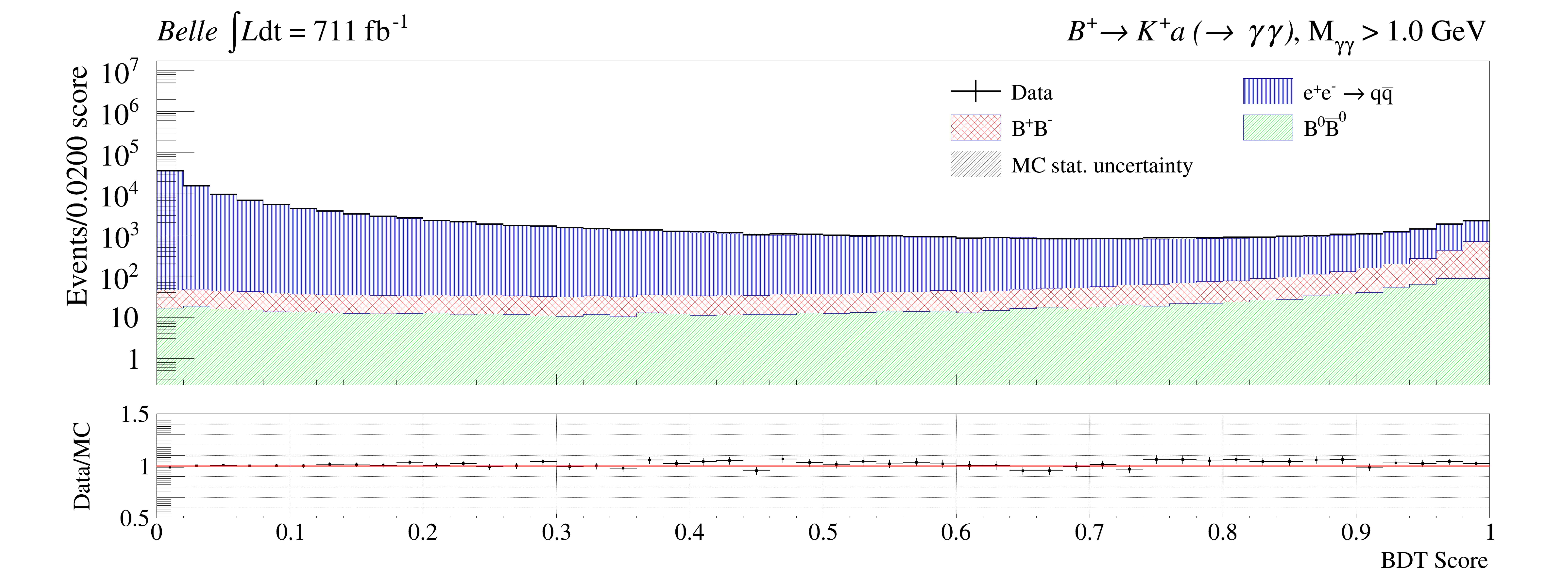}
\caption{\label{fig:BDT-b} 
Distributions of the second continuum suppression BDT classifier scores.
The colour convention used in this histogram is identical to that in Fig.~\ref{fig:BDT-a}}
\end{figure}

\begin{figure}[ht]
\includegraphics[width=1.0\linewidth]{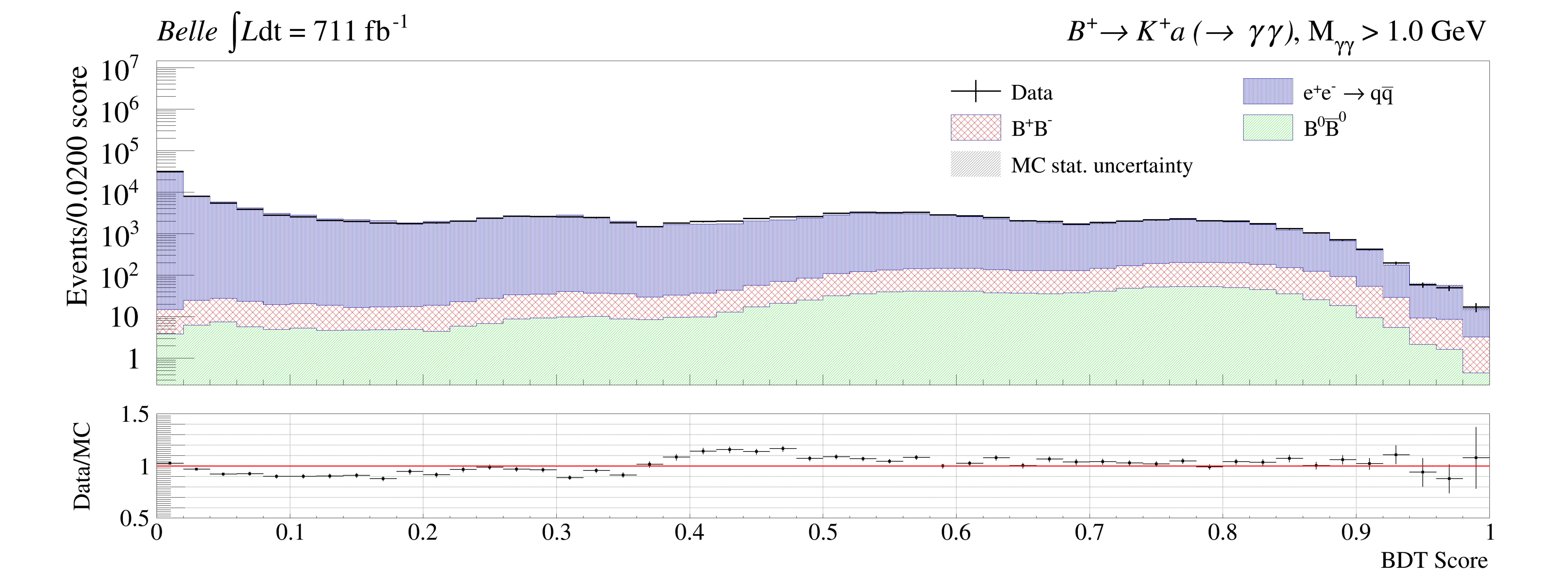}
\caption{\label{fig:BDT-c} 
Distributions of $P_{\piz}(\g_1)$ BDT classifier scores.
The colour convention used in this histogram is identical to that in Fig.~\ref{fig:BDT-a}}
\end{figure}

\begin{figure}[ht]
\includegraphics[width=1.0\linewidth]{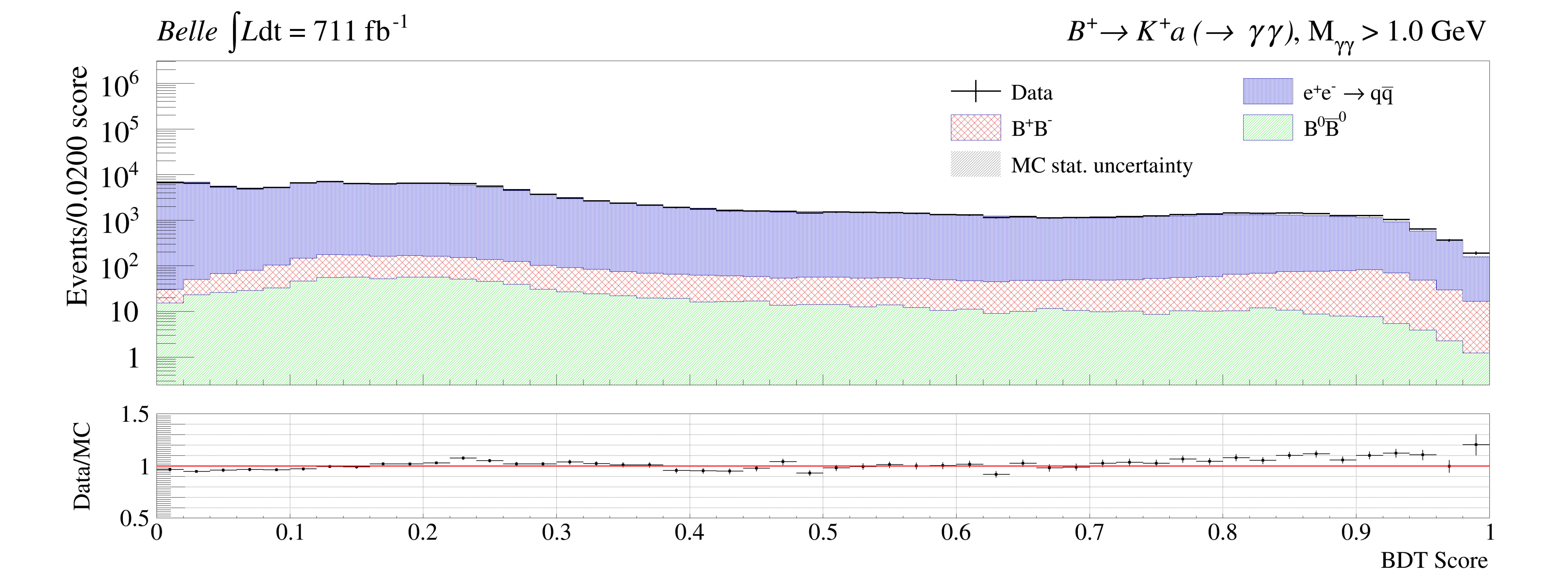}
\caption{\label{fig:BDT-d} 
Distributions of $P_{\piz}(\g_2)$BDT classifier scores.
The colour convention used in this histogram is identical to that in Fig.~\ref{fig:BDT-a}}
\end{figure}

\begin{figure}[ht]
\includegraphics[width=1.0\linewidth]{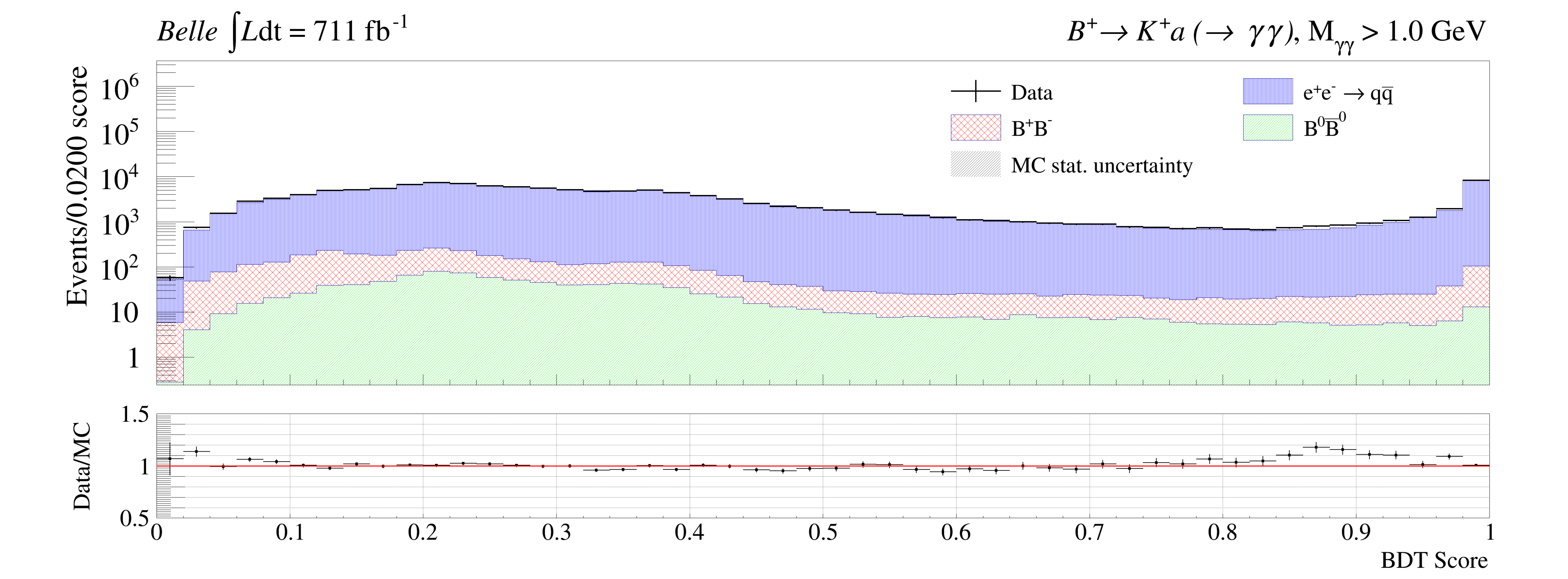}
\caption{\label{fig:BDT-e} 
Distributions of $\Xsg$ suppression BDT classifier scores.
The colour convention used in this histogram is identical to that in Fig.~\ref{fig:BDT-a}}
\end{figure}

\begin{figure}[ht]
\subfloat[\label{fig:Mgg_data-00}]{\includegraphics[width=1.0\linewidth]{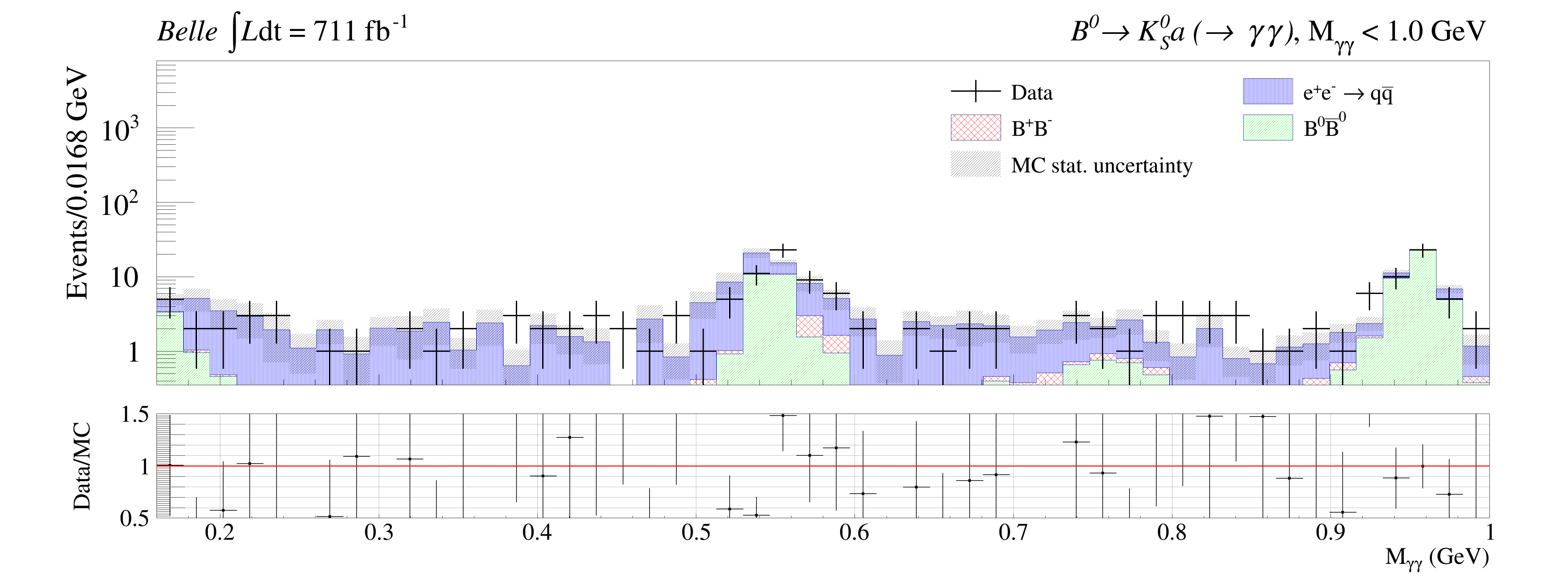}}\hfill
\subfloat[\label{fig:Mgg_data-01}]{\includegraphics[width=1.0\linewidth]{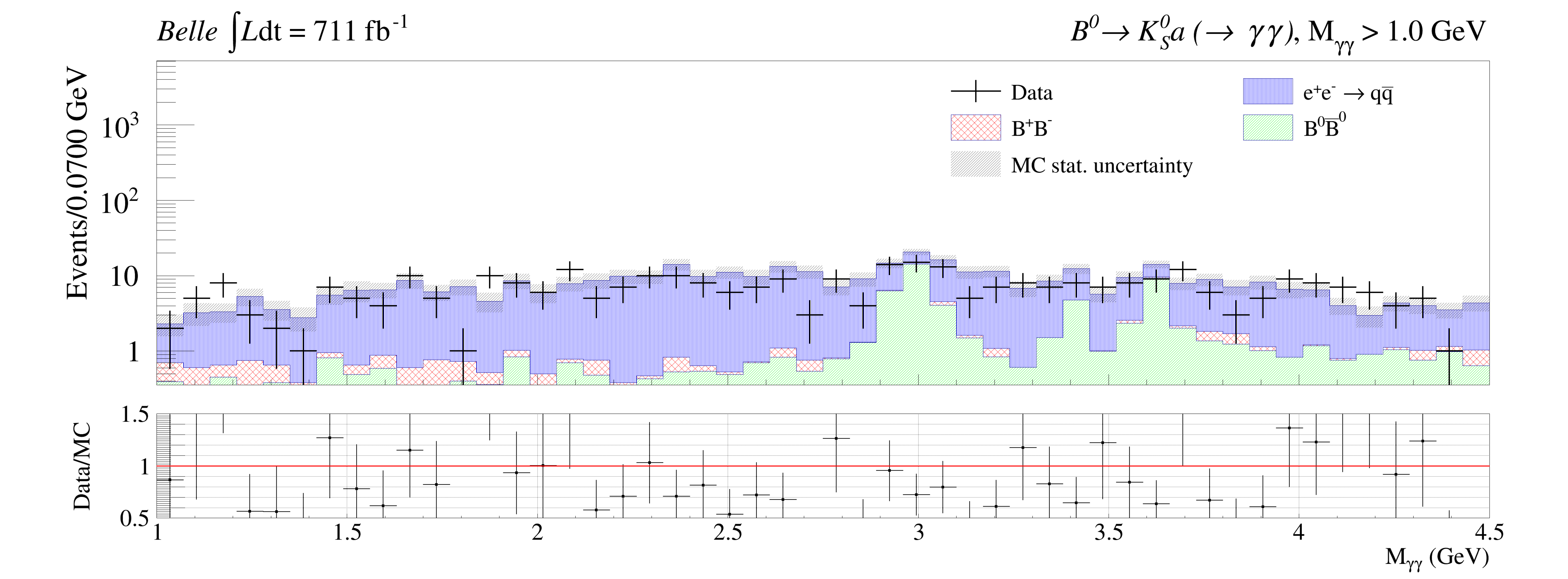}}\hfill
\caption{\label{fig:Mgg-sup0} 
Diphoton invariant mass distribution of ALP candidates in $\Bz\ra\KS\ap$ decay,
The colour convention used in this histogram is identical to that in Fig.~\ref{fig:Mgg}}
\end{figure}

\begin{figure}[ht]
\subfloat[\label{fig:Mgg_data-20}]{\includegraphics[width=1.0\linewidth]{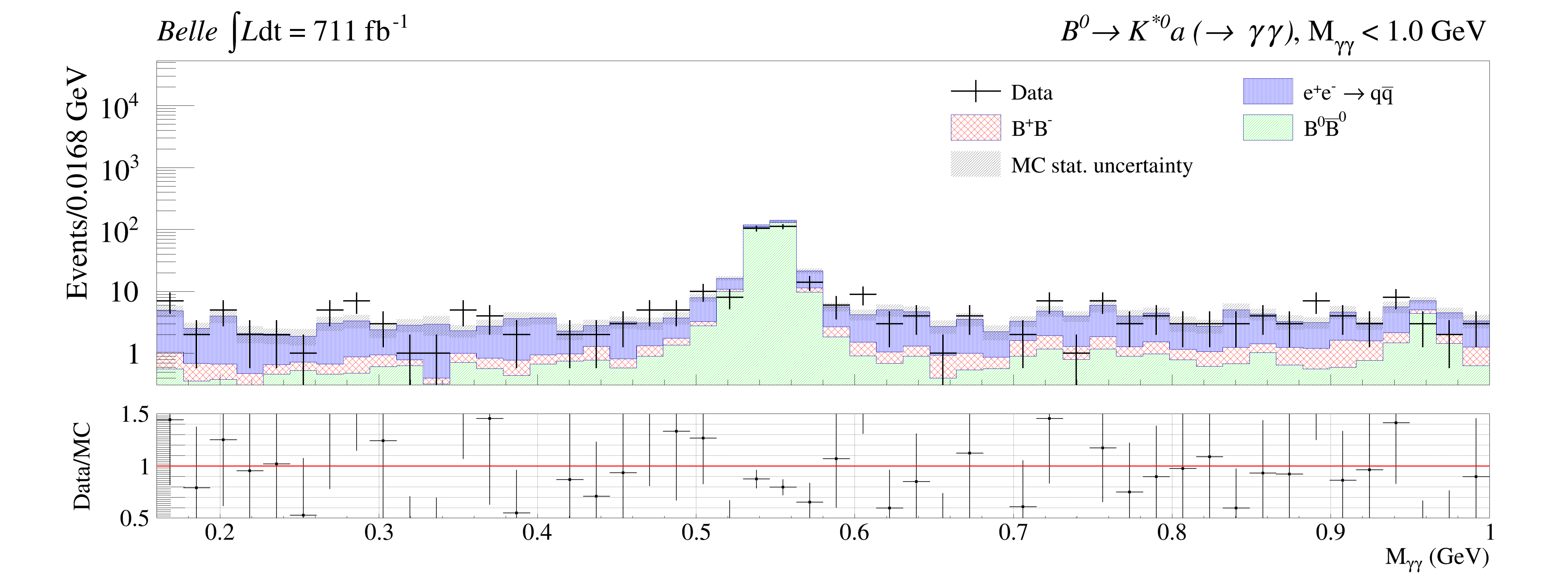}}\hfill
\subfloat[\label{fig:Mgg_data-21}]{\includegraphics[width=1.0\linewidth]{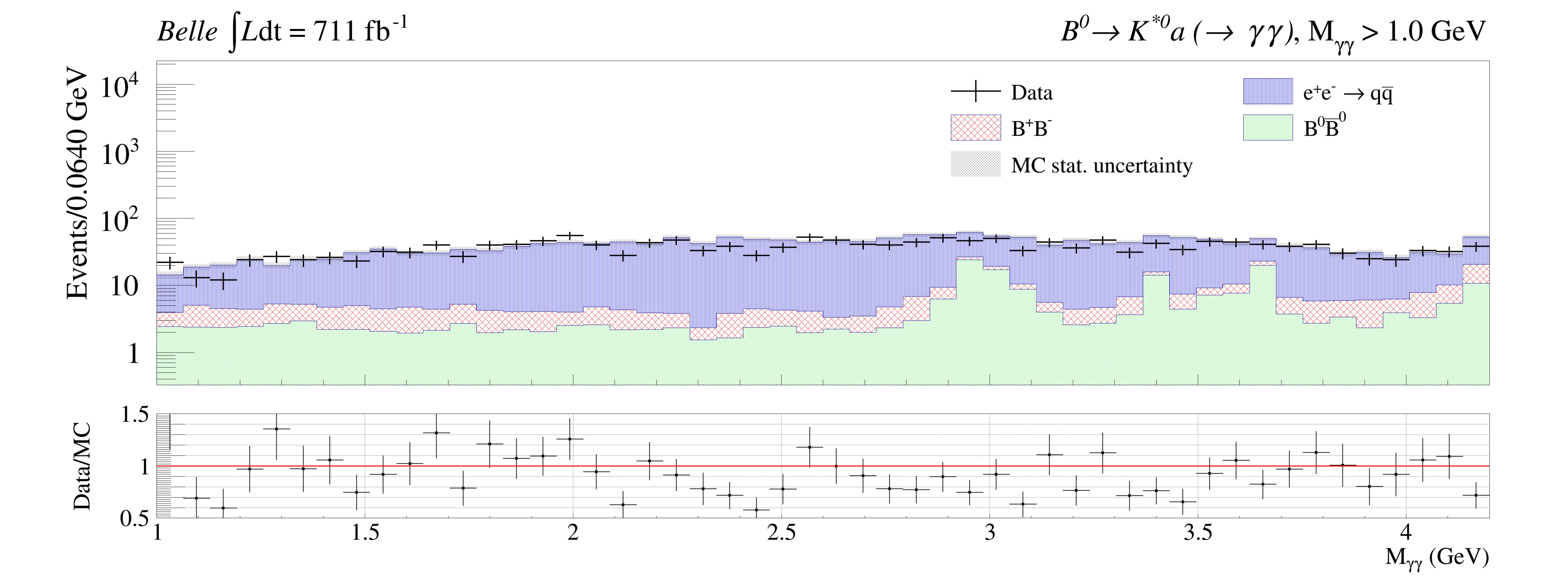}}\hfill
\caption{\label{fig:Mgg-sup2} 
Diphoton invariant mass distribution of ALP candidates in $\Bz\ra\Kstarz\ap$ decay,
The colour convention used in this histogram is identical to that in Fig.~\ref{fig:Mgg}}
\end{figure}   

\begin{figure}[ht]
\subfloat[\label{fig:Mgg_data-30}]{\includegraphics[width=1.0\linewidth]{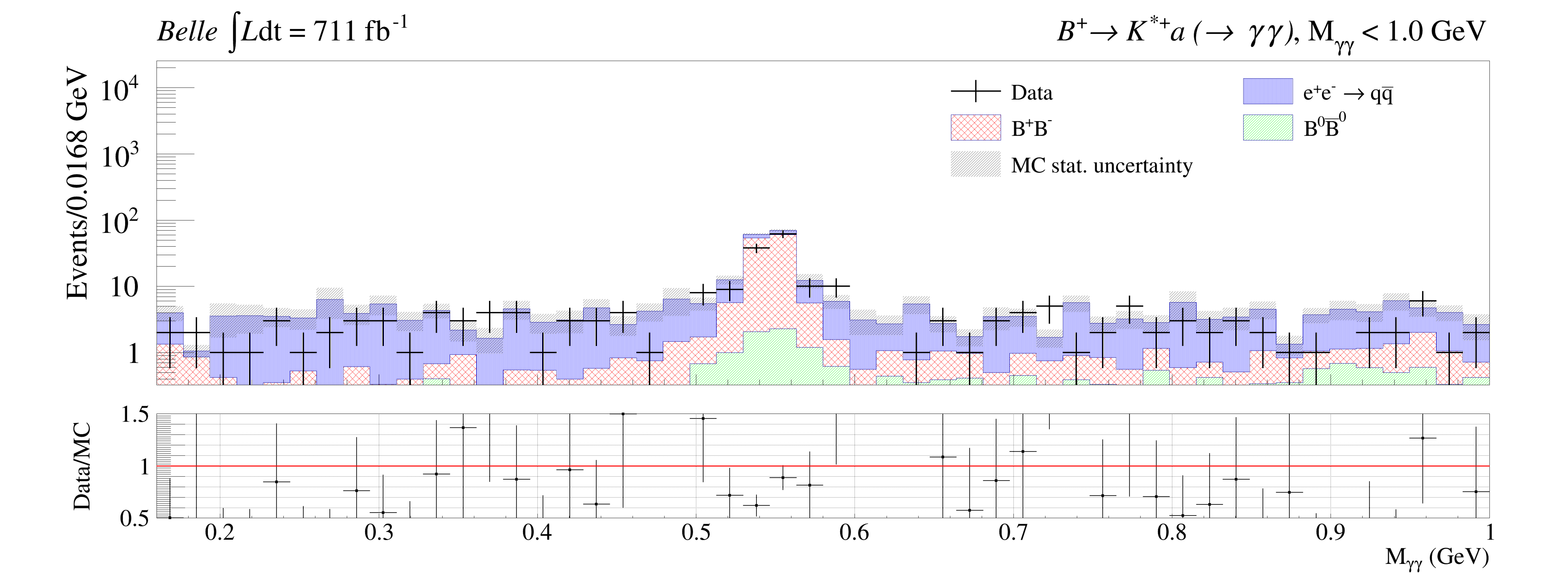}}\hfill
\subfloat[\label{fig:Mgg_data-31}]{\includegraphics[width=1.0\linewidth]{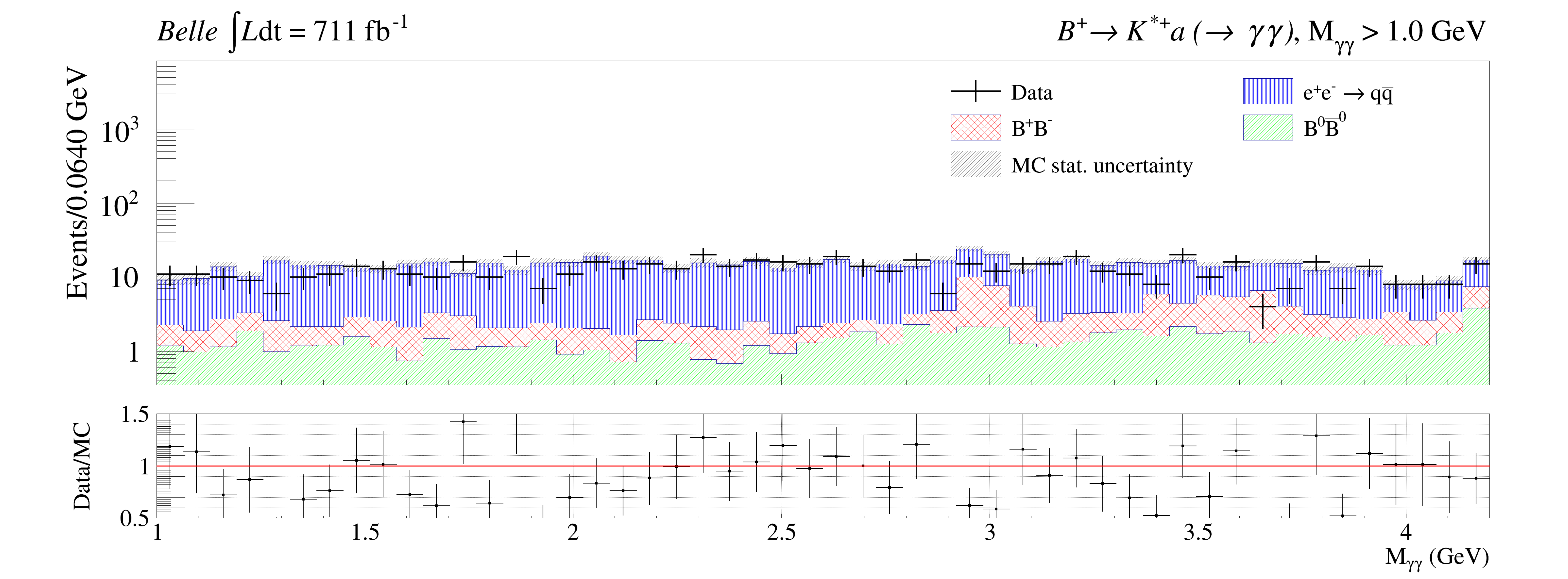}}\hfill
\caption{\label{fig:Mgg-sup3} 
Diphoton invariant mass distribution of ALP candidates in $\Bp\ra\Kstarp\ap$ decay,
The colour convention used in this histogram is identical to that in Fig.~\ref{fig:Mgg}}
\end{figure}   

 \begin{figure}[ht]
 \includegraphics[width=1.0\linewidth]{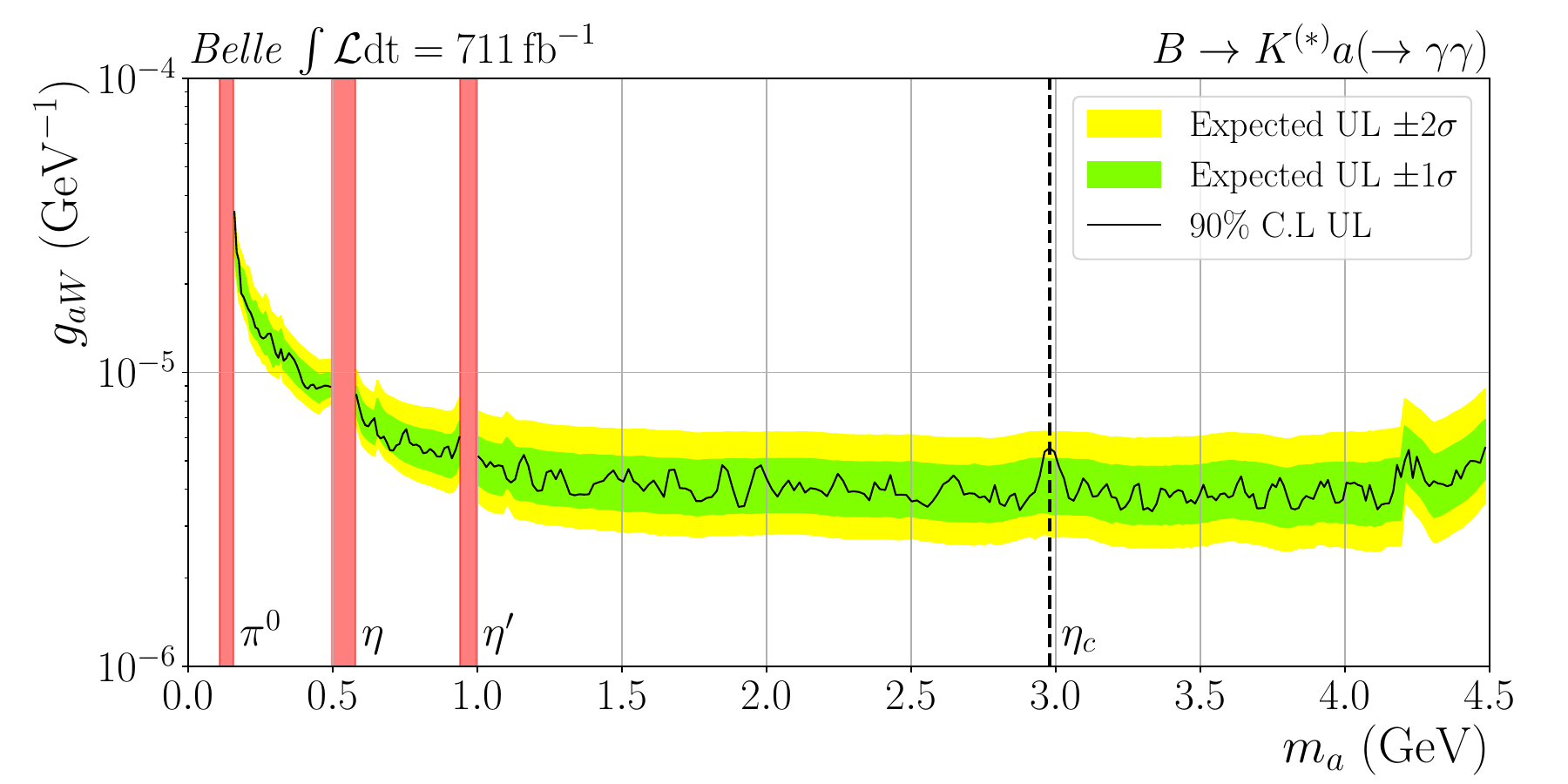}
 \caption{\label{fig:gaWUL-sup} 
 The 95\% CL upper limits on the coupling $\gaW$ as a function of the ALP mass obtained with CLs method with simultaneous fit. 
 The green and yellow bands are $\pm1$ and $\pm2$ standard deviation ranges, respectively, of expected upper limit of background only model. 
 The red bands are the excluded $\piz,\eta$ and $\etp$ mass regions.
 The vertical dashed line indicates the nominal $\etc$ mass.}
 \end{figure}

\begin{figure}[ht]
\includegraphics[width=1.0\linewidth]{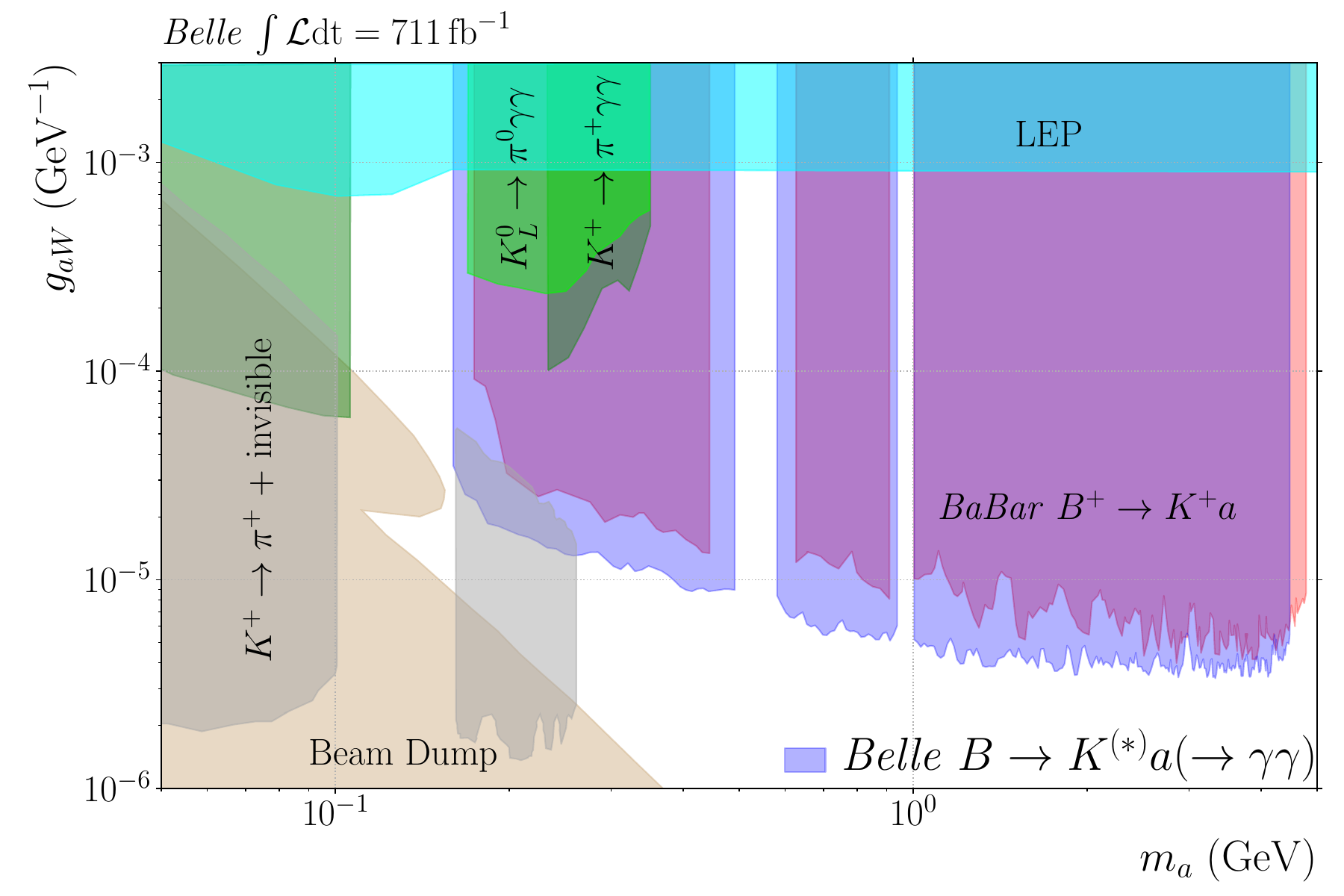}
\caption{\label{fig:gaWExc-sup} 
The 95\% CL upper limits on the coupling $\gaW$ 
from a simultaneous fit to the four $\B\to\kaon^{(*)}\ap$ modes as a function of the ALP mass,
compared with existing constraints.}
\end{figure}

\end{document}